\documentclass[sigconf, nonacm]{acmart}
\usepackage{algorithm}
\usepackage{algpseudocode}
\usepackage{tabularx}
\usepackage{xcolor}
\usepackage{bbold}

\usepackage{enumitem}
\usepackage{multirow}
\usepackage{marginnote}
\usepackage{balance}
\usepackage{multicol} 
\usepackage{lipsum}   
\usepackage{array}    

\usepackage{xcolor}
\definecolor{darkgreen}{RGB}{0,180,0}
\newcommand{\lzdh}[1]{\textcolor{black}{#1}}
\definecolor{l_green}{RGB}{136, 182, 181}
\newcommand{\sjc}[1]{\textcolor{black}{#1}}

\begin{document}
\title{AMAZe: A Multi-Agent Zero-shot Index Advisor for Relational Databases}

\author{Zhaodonghui Li}
\orcid{0009-0009-8409-3374}
\affiliation{%
  \institution{Nanyang Technology University}
}
\email{G220002@e.ntu.edu.sg}

\author{Haitao Yuan}
\orcid{0000-0002-1825-0097}
\affiliation{%
  \institution{Nanyang Technology University}
}
\email{haitao.yuan@ntu.edu.sg}

\author{Jiachen Shi}
\orcid{0000-0002-1825-0097}
\affiliation{%
  \institution{Nanyang Technology University}
}
\email{Jiachen001@e.ntu.edu.sg}

\author{Hao Zhang}
\orcid{0000-0002-2725-6458}
\affiliation{%
  \institution{DAMO Academy, Alibaba Group}
}
\email{hzhang26@outlook.com}

\author{Yu Rong}
\orcid{0000-0001-7387-302X}
\affiliation{%
  \institution{DAMO Academy, Alibaba Group}
  \institution{Hupan Lab, Hangzhou, China}
}
\email{yu.rong@hotmail.com}

\author{Gao Cong}
\orcid{0000-0002-1825-0097}
\affiliation{%
  \institution{Nanyang Technology University}
}
\email{gaocong@ntu.edu.sg}

\begin{abstract}
Index recommendation is one of the most important problems in database management system (DBMS) optimization. Given queries and certain index-related constraints, traditional methods rely on heuristic optimization or learning-based models to select effective indexes and improve query performance. However, heuristic optimization suffers from high computation time, and learning-based models lose generalisability due to training for different workloads and database schemas. With the recent rapid development of large language models (LLMs), methods using prompt tuning have been proposed to enhance the efficiency of index selection. However, such methods still can not achieve the state-of-the-art (SOTA) results, and preparing the index selection demonstrations is also resource-intensive. To address these issues, we propose AMAZe, a zero-shot LLM-based index advisor with a multi-agent framework. We decompose the index recommendation problem into sub-steps, including planning, selection, combination, revision, and reflection. A set of LLM-embedded agents is designed to handle each one of the different sub-steps. Our method utilizes high-level agents to control the index selection process and low-level agents to select and revise indexes. Through extensive experiments, we show that our proposed AMAZe not only achieves the SOTA performance compared to the heuristic methods, but also outperforms learning-based and prompt-based methods with higher efficiency and better zero-shot inference ability.
\end{abstract}

\maketitle

\section{Introduction}
Indexes in database systems significantly accelerate data retrieval and reduce redundant I/O operations through providing alternative access paths to data based on attribute values.
With the appropriate index configuration, database systems can better support analytical workloads that are highly demanded by many modern applications such as inventory management, online shopping and real-time advertisement.
However, selecting and creating indexes always comes with significant computation time and storage cost. Therefore, given the constraints such as limited resource and budget, it is important for database administrators (DBAs) to select the optimal indexes that can meet the data management needs.

\begin{table*}[htp]
\caption{Overview of the existing Index Advisors, evaluated in terms of the three fundamental challenges.}
\label{tab:advisors}
\scalebox{0.9}{
\begin{tabular}{|l|l|l|l|l|} \hline
             & Heuristic-based method & Learning-based method & Prompt-based method & AMAZe (Ours)  \\ \hline
Zero-shot Generalization 
 & Yes & No & No & Yes     \\ \hline 
Effectiveness  & High & Medium & Medium & High   \\ \hline
Efficiency  & Low & Low & High & High     \\ \hline 
\end{tabular}}
\end{table*}

To assist 
DBAs with index selection,
index advisors~\cite{chaudhuri2020anytime, 10.1145/1066157.1066184, 10.1145/3340531.3412106, DBLP:conf/icde/PereraORB21, DRLindex, 9094124, 10.5555/846219.847390, balance, extend, 315323, 10.5555/645923.673646, swirl, llmidxadvis, autoindex2022, 10.1007/s10489-020-01674-8, DBAbandits, budgetaware, ia2, mfix} are often adopted to automatically select the optimal indexes for a given query workload in the target database. Existing index advisors fall into three main categories: heuristic-based advisors~\cite{chaudhuri2020anytime, 10.1145/1066157.1066184, 10.5555/846219.847390, extend, 315323, 10.5555/645923.673646}, learning-based advisors~\cite{10.1145/3340531.3412106, DBLP:conf/icde/PereraORB21, DRLindex, 9094124, balance, swirl, autoindex2022, 10.1007/s10489-020-01674-8, DBAbandits, budgetaware, ia2, mfix}, and prompt-based LLM advisors~\cite{llmidxadvis}. 
\textbf{Heuristic-based advisors} typically use rule-based or greedy algorithms to identify optimal indexes based on database optimizers' cost estimations. However, given the large number of columns involved in complex workloads, exploring exhaustively the search space for all possible index combinations becomes prohibitory expensive. In addition, the reliance on heuristics and optimizers' estimations can lead to suboptimal index selection. As highlighted in~\cite{aimeetsai}, the errors in cost estimations can even cause workload performance degradation, a phenomenon known as the index regression problem.
\textbf{Learning-based advisors}
have emerged as a promising alternative to traditional heuristic-based approaches. They
utilize techniques like deep learning and reinforcement learning to train index selection models and index benefit estimators. 
Through extensive learning on features extracted from queries, execution plans, and database statistics, these models can capture complex interactions between indexes and queries that are difficult to encode manually and hence achieve SOTA performance. However, training these models often requires a large amount of labeled data and time. The lack of ability to generalize to unseen database limits their applicability in dynamic or evolving environments. Moreover, the need for data annotations and model re-training poses a major limitation in data-sensitive domains like healthcare or security, where data collection and sharing are tightly regulated.
To exploit the adaptability and semantic reasoning strengths of LLMs while mitigating both database optimizer dependency and data dependency,
\textbf{prompt tuning on LLMs}~\cite{zhou2023large} is adopted. Such a method~\cite{llmidxadvis} prompts LLMs with preprocessed index recommendation demonstrations and 
allows the LLM select columns to build the indexes. 
While this LLM-based approach enhances the generalization and efficiency of index advisors, it also presents notable limitations.
Actually, prompt tuning approach exhibits limited effectiveness in zero-shot setting due to LLMs' limited understanding of database-specific contexts and optimization nuances without concrete examples, necessitating few-shot learning with domain-specific demonstrations~\cite{brown2020language}. However, preparing high-quality demonstrations is challenging and time-consuming. In addition,  inference with LLMs can be computationally expensive for large-scale workloads.
When processing complex workloads, the number of tokens to be encoded and reasoned over grows rapidly, leading to high memory consumption and long inference times. This computational overhead becomes particularly pronounced in online environments.
Table~\ref{tab:advisors} summarizes the characteristics and drawbacks of three types of advisors as discussed above. 

Given the limitations of existing methods, we aim to design an index advisor that achieves the following key objectives in index recommendation:
(1) \textbf{Zero-shot Generalization}: To accommodate the diverse landscape of real-world database deployments, an effective index advisor must exhibit strong generalization capabilities. Specifically, the ability to recommend optimal indexes for previously unseen databases and workloads without requiring retraining or fine-tuning on domain-specific data.
(2) \textbf{Effectiveness}: The advisor should deliver SOTA recommendation quality, producing index configurations that demonstrably improve query performance compared to existing approaches. (3) \textbf{Efficiency}: The advisor must exhibit low computational overhead, enabling rapid index recommendations that scale to large workloads without imposing significant time costs. This efficiency is critical because database systems often handle massive datasets and complex query patterns where slow workload processing and index recommendation can delay essential optimizations, potentially impacting overall system performance and user experience.

In order to achieve the objectives above,
we propose AMAZe, \underline{a} \underline{m}ulti-\underline{a}gent \underline{z}ero-shot ind\underline{e}x advisor based on LLM. 
\sjc{To equip AMAZe with zero-shot inference capability,}
we 
build upon 
\sjc{recent advances in multi-agent frameworks that have demonstrated success in addressing other tasks}
under the zero-shot setting~\cite{liu2025towards, li2025agentoriented, jha2025crossenvironment}. 
\sjc{A multi-agent framework can better support zero-shot inference because it decomposes the complex index selection problem into specialized sub-tasks, allowing different agents to collaborate and compensate for the limitations of a single LLM. These sub-tasks enable the framework to leverage the general knowledge encoded in LLMs while structuring their reasoning in a way that aligns more closely with database-specific requirements.}
\lzdh{We further design three critical components within the multi-agent framework: \textbf{efficient workload representation}, \textbf{budget-aware agent planning}, and \textbf{effectiveness-aware agent coordination}. The efficient workload representation facilitates streamlined communication among agents and mitigates LLM hallucination by providing a concise yet informative summary of the workload. Complementarily, budget-aware agent planning enhances overall efficiency by guiding the LLM to select indexes that optimize resource usage while respecting storage and computation constraints. To tackle the index regression problem, which constitutes the primary barrier to achieving effectiveness, we design effectiveness-aware agent coordination.
Through cooperative interactions, multiple agents iteratively refine the current index selection by evaluating index benefits, thereby ensuring a more robust, stable, and effective index recommendation pipeline.}

We conduct extensive experiments and recommend indexes for multiple popular datasets and workloads, including TPC-H, TPC-DS, DSB, and JOB. We observe that AMAZe's benefit-to-cost ratio outperforms all the heuristic-based baselines, which are also zero-shot, in 10 out of the 16 settings, 
and matches the best of them in the remaining 6 settings.
In addition, AMAZe significantly outperforms the learning-based methods which requires intensive training. Compared with the few-shots prompt-based baseline, AMAZe also achieves better performance with lower inference time without relying on domain-specific data.

Our main contributions are as follows:
\vspace{-\topsep}
\begin{itemize}[leftmargin=*]
  \item We propose AMAZe, which achieves SOTA results with higher efficiency and zero-shot generalization. To the best of our knowledge, AMAZe is the first LLM-based zero-shot index advisor.
  \item We propose a new workload representation paradigm that is suitable for LLMs to understand and analyze the index recommendation problem. 
  \item We design a multi-agent framework to solve the index recommendation problem. The problem is decomposed into multiple sub-steps, which are much easier to solve by a cooperation of the high-level and low-level agents. 
  \item Extensive experiments on various workloads demonstrate that AMAZe achieves superior performance with lower inference time compared to baseline methods.

\end{itemize}
\vspace{-\topsep}

\section{Preliminary}
In this section, we will first clarify the concepts related to index recommendation in Section~\ref{sec:concept}. Next, we will then formally define the problem of an index advisor in Section~\ref{sec:index}.

\subsection{Workload, Index Candidate and Constraints~\label{sec:concept}}
\textbf{Workload}: The workload is referred to as a set of queries that will be 
\sjc{optimized} in the database. The ultimate goal of the index recommendation problem is to optimize the execution efficiency of the entire workload instead of any specific query within it.

\noindent\textbf{Index Candidate}: Most of the methods that solve the index recommendation problem start with collecting a set of index candidates to select from. Each candidate within this set usually stands for a table column 
\sjc{accessed}
by any of the queries in the workload.

\noindent\textbf{\sjc{Selection Constraints}}: Index selection in databases is subject to various practical constraints, \sjc{such as index numbers and} storage budget. 
\sjc{Selection constraints in the paper is referred as the additional storage overhead (i.e., disk space) unless otherwise specified.} During index selection, it is significant to balance performance gains against the increased resource usage, which necessitates prioritizing indexes that offer the highest benefit-to-cost ratio.

\subsection{Index Advisor~\label{sec:index}}
Index advisors receive the workload of a specified database along with \sjc{selection} constraints, 
and then automatically generate \sjc{a set of recommended indexes.}
The functionality of an index advisor can be formally defined as follows:

\begin{definition}
(Index Advisor): Given a database $D$, a workload $W = \{q_1, q_2, \ldots, q_n\}$ consisting of a set of queries, and a storage budget $B$, the index advisor problem seeks to automatically identify an optimal set of indexes $I$ that minimizes workload execution cost while respecting storage constraints. Formally, let $C$ denote the set of all possible column combinations that can be indexed in database $D$. The index advisor solves the following optimization problem:
\begin{displaymath}
  I^* = \arg\!\min_{I \subseteq \mathcal{P}(C)} Cost(D, W, I) \quad \text{s.t.} \quad Storage(I) \leq B
\end{displaymath}
where $\mathcal{P}(C)$ represents the power set of all possible column combinations, $Cost(D, W, I)$ computes the total execution time of workload $W$ on database $D$ with index set $I$,
and $Storage(I)$ calculates the total storage cost to construct the indexes $I$ selected.
\end{definition}

\section{Related Work}
\subsection{Heuristic-based Index Advisors}
Heuristic-based advisors iteratively select candidate indexes for an input workload in a greedy fashion. Within each optimization step, the index advisors will update the index recommendation according to the estimated benefits of the index candidates. There are mainly two heuristics for the index advisors to follow: ADD heuristic and DROP heuristic. As an example of ADD heuristic, Extend~\cite{extend} starts with no indexes and add one index at a time according to the highest benefit-to-cost ratio \lzdh{until the budget constraint is met}. In contrast, Drop~\cite{315323} following the DROP heuristic starts with all index candidates and iteratively removes the least beneficial one until \lzdh{only the desired number of indexes are left}. In general, heuristic-based advisors generally take significantly longer computation time to recommend indexes, especially for databases with complicated schemas. In addition, heuristic-based methods rely solely on database optimizers to evaluate index benefits, and 
wrong estimations in column cardinality may mislead index advisors to select indexes that are not beneficial to the workload~\cite{aimeetsai}.


\subsection{Learning-based Index Advisors} 
State-of-the-art learning-based advisors~\cite{swirl, balance, 10.1145/3340531.3412106, DRLindex, 10.1007/s10489-020-01674-8, budgetaware, ia2} predominantly adopt reinforcement learning (RL) algorithms for lower inference latency. For example, DQN~\cite{10.1145/3340531.3412106} integrates Q-learning with deep neural networks to derive optimal policies from workload representations, while SWIRL~\cite{swirl} and BALANCE~\cite{balance} leverage the PPO~\cite{ppo} algorithm for policy optimization, with BALANCE further employing a transfer learning framework to exploit historical workloads. However, RL-based methods need significantly longer training time and have lower generalisability, as training is required for each unseen database. \lzdh{Alternative learned models are proposed to address the limitations. For example, DBA Bandits~\cite{DBAbandits} employs a Multi-Armed Bandits framework, where index selection policies are defined and refined through neural networks to enhance computational efficiency. AutoIndex~\cite{autoindex2022} leverages a Monte Carlo Tree Search algorithm, adopting an ADD-style strategy similar to heuristic-based methods for index selection. More recently, MFIX~\cite{mfix} introduces a compact tree-structured index space and improves search efficiency by utilizing a multi-fidelity Bayesian optimization (MFBO) algorithm. Despite these advances, such methods often suffer from poor generalization or trade-offs between efficiency and effectiveness. Moreover, they still require retraining when the underlying data changes and fail to operate in zero-shot scenarios. It is noteworthy that some recent work focuses on specific sub-tasks of index recommendation including workload compressing~\cite{isum, wred}, index candidate filtering~\cite{distill} and index evaluation~\cite{queryformer}. As these studies require considerable extra design to establish a complete index advisor, we leave them for future discussion.}

\subsection{Prompt-based Index Advisors}
\sjc{Recently, LLMIdxAdvis~\cite{llmidxadvis}  first utilizes LLMs to generate index suggestions. Specifically, it leverages a prepared demonstration pool and employs cosine similarity to select the most relevant in-context demonstrations for a given workload. While this strategy improves efficiency and generalizability in index selection, these gains come at the expense of recommendation quality. Specifically, LLMIdxAdvis struggles to handle complex workloads and database schemas due to its reliance on limited demonstrations. In addition, it progressively builds index in a single-agent, greedy manner without refinement or verification, leading to significant hallucination problems. Deviating from LLMIdxAdvis, our work explores a multi-agent framework, which doesn't rely on demonstrations and mitigates hallucinations via cross-agent coordination.}

\section{Framework Overview}

In this section, we will give an overview of how we design our AMAZe framework to address the current challenges for index recommendation problem. In Section~\ref{sec:41}, we will first explain how we solve the zero-shot generalization challenge using a multi-agent pipeline. In Section~\ref{sec:42} we will then 
introduce our novel framework designs, namely the workload representation paradigm and effectiveness-aware agents for better effectiveness and efficiency.




\begin{table}[t]
\caption{
Preliminary experiment on 100 random TPC-DS workloads under 2 GB storage budget. The mean execution time of the 100 workloads with no index, with Extend's recommendation, and GPT-4o's recommendation is listed.}
\label{tab:pioneer}
\begin{tabular}{|l|l|l|l|l|} \hline
Execution time (s)   &  With regressions & Remove regressions \\ \hline
No index  & 193.58
  & 193.58   \\ \hline
Extend   & \textbf{161.28}
   & 160.53
    \\ \hline 
GPT-4o  & 186.70
  & \textbf{155.54}
 \\ \hline 
\end{tabular}
\end{table}

\begin{figure*}[t]
  \centering
  \includegraphics[width=\linewidth]{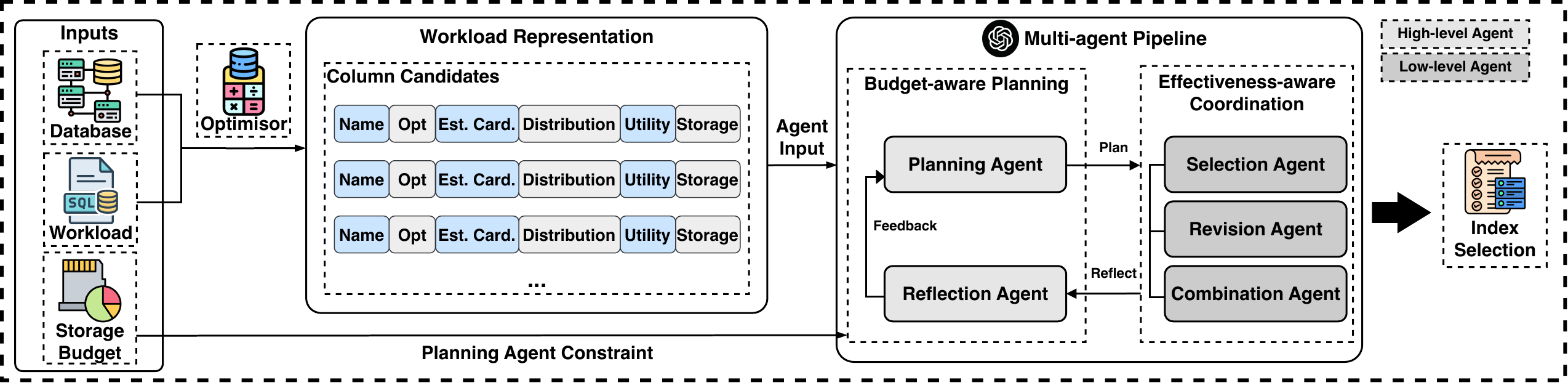}
  \caption{The overview of AMAZe, the zero-shot multi-agent index advisor consisting of workload representation (Section~\ref{sec:41}) and a multi-agent pipeline (Section~\ref{sec:42}).}
  \Description{AMAZe}
  \label{fig:pipeline}
\end{figure*}

\subsection{Multi-agent Based Pipeline}
\label{sec:41}

\noindent \textbf{Motivation.} Although LLMs demonstrate strong potential for zero-shot generalization in index recommendation, it is hard to optimize the LLM prompts for each database schema. To investigate this limitation, we conduct a preliminary experiment on the TPC-DS benchmark. We construct 100 small workloads, each containing 3–10 randomly selected queries over the TPC-DS schema, under a fixed 2 GB storage budget. We compare the performance of GPT-4o~\cite{openai2023gpt4}, applied directly in a zero-shot manner with workload and query plan inputs, against the heuristic baseline Extend~\cite{extend}. Table~\ref{tab:pioneer} shows that GPT-4o underperforms Extend by an average of 15.8 \% (25.42 seconds). While GPT-4o 
can generate reasonable index recommendations without additional training, its performance remains limited and inconsistent across complex workloads.
These findings confirm that \textbf{index recommendation is far too complex to be solved effectively with a single zero-shot LLM call}, highlighting the need for a \textbf{well-designed LLM-powered framework} that can fully harness LLMs' zero-shot generalization capabilities beyond direct prompting alone.

Inspired by the recent advancements in multi-agent LLM pipelines that divide a complex problem into sub-problems handled by specialized agents~\cite{liu2025towards, li2025agentoriented, jha2025crossenvironment}, we aim to decompose the index recommendation task into multiple sub-tasks like index selection and combination. This design enhances scalability, interpretability, and efficiency, particularly well-suited for complex workloads that demand coordination, specialization, and iterative refinement.

\noindent\textbf{Pipeline Overview.} We introduce a solution based on \textbf{a hierarchical multi-agent based index recommendation pipeline}, illustrated in Figure~\ref{fig:pipeline}. At the high-level, the pipeline reflectively decomposes the complex index recommendation task into a sequence of LLM-solvable sub-steps, such as selecting promising column candidates, revising suboptimal or redundant choices, and composing indexes when beneficial. At the low-level, each sub-step is handled by a specialized agent powered by an LLM, ensuring that decisions remain accurate and contextually appropriate. This design directly addresses the challenge that LLMs struggle with the full index recommendation problem in a single zero-shot call: the multi-agent pipeline excels at solving simpler, well-defined tasks with greater accuracy and consistency~\cite{brown2020language}. By orchestrating multiple agents, our pipeline leverages the generalization power of LLMs while avoiding their weaknesses on overly complex reasoning. 


\subsection{Key Component Designs}
\label{sec:42}
\noindent \textbf{Design Rationale.} To develop an effective multi-agent framework, we must solve three critical design problems: (1) how to efficiently represent workloads for LLM agents, (2) how to incorporate storage budget constraints into the agent pipeline, and (3) how to prevent index regression that undermines recommendation quality. These design challenges are elaborated below:

\noindent \underline{(1) Workload Representation.} Jointly encoding database schemas, workloads, and query execution plans often exceeds the context lenght limitations of LLMs—the TPC-DS workload with 83 queries requires over 36,000 tokens, surpassing most LLM's context capacity and incurring substantial costs. Beyond computational constraints, long contexts reduce agent scalability and decision accuracy.
Moreover, existing workload representations are unsuitable for agent-based systems. Learning-based methods encode information into embedded features~\cite{10.1145/3340531.3412106, balance} that agents cannot interpret, while prompt-based approaches~\cite{llmidxadvis} encode each query into a nested dictionary and append them as a list input that agents struggle to navigate effectively. These verbose, code-format representations misalign with how LLM agents process and reason about information. \textbf{Therefore, we need to design a specialized workload representation technique that enables efficient agent communication and reasoning.}

\noindent \underline{(2) Budget Constraints.} Existing agent frameworks lack mechanisms to enforce storage constraints across multiple agents. Our preliminary experiments reveal that LLM does not always know when to stop the recommendation, demonstrating that agents operating without budget coordination produce inconsistent and resource-inefficient results. 
\textbf{Thus, we must design budget-aware coordination mechanisms that enable agents to collaborate while respecting storage limitations.}

\noindent \underline{(3) Regression Prevention.} The index regression problem—where recommended indexes degrade query performance-is amplified in multi-agent settings due to three factors: suboptimal decisions by individual agents, LLM hallucinations~\cite{Ji_2023, zhang2023sirens}, and compounding errors across agent interactions. As shown in Table~\ref{tab:pioneer}, when regressing indexes are removed, GPT-4o surpasses traditional baselines, indicating that not only regression prevention is crucial, but also optimal index recommendation is achievable by LLMs. However, general-purpose LLMs lack database domain knowledge~\cite{wei2023larger}, leading agents to make uninformed decisions that cascade through the pipeline. Existing agent frameworks provide no mechanisms to detect, prevent, or correct such regression patterns. \textbf{Consequently, we need to design agent coordination strategies that incorporate database-specific validation, decompose complex decisions into reliable subtasks, and implement feedback mechanisms to prevent regression propagation.}

\smallskip

\noindent \textbf{Key designs.}  To address the aforementioned challenges, we propose three key designs that enable effective multi-agent coordination for index recommendation.

\noindent \underline{(1) Efficient Workload Representation.} We design a novel textual workload representation that transforms the complex index recommendation problem into an LLM-compatible format. As illustrated in Figure~\ref{fig:pipeline}, our approach extracts only index-relevant information and reorganizes it at the column level, merging all workload details associated with each database column to construct a compact set of indexable candidates. This design provides two critical advantages. First, it ensures efficiency and scalability by bounding the input length to the number of database columns rather than the number of queries—making the representation size independent of workload complexity. Second, it reduces hallucination by reframing index recommendation as a structured selection task where agents choose and combine predefined indexable columns, aligning the problem with LLMs' decision-making strengths rather than requiring free-form generation.

\noindent \underline{(2) Budget-Aware Agent Planning.} Our pipeline incorporates budget-aware coordination mechanisms that enable high-level agents to make resource-conscious decisions throughout the recommendation process. The planning agent receives the current storage consumption at its stage and communicates constraints to subsequent low-level agents, ensuring that the final recommendations remain feasible under real-world storage limitations. This index budget management prevents inefficient computations while maintaining the flexibility for agents to explore diverse optimization strategies within acceptable bounds.

\noindent \underline{(3) Effectiveness-Aware Agent Coordination.} To prevent index regression, we design specialized low-level agents targeting the two primary causes of performance degradation: LLM hallucination and database optimizer estimation errors. As shown in Figure~\ref{fig:pipeline}, our coordination strategy employs two complementary agents: \textbf{The combination agent} addresses optimizer estimation bias by merging individually suboptimal indexes into composite indexes that better capture multi-column access patterns, effectively combining inefficient recommendations into more beneficial ones. \textbf{The revision agent} eliminates LLM-generated errors, including hallucinated indexes that do not exist in the schema, redundant duplicates, and indexes that empirically degrade performance. 
Importantly, the cooperation between these two agents not only addresses hallucination and misestimation but also simulates \textbf{a hybrid of ADD and DROP index recommendation heuristics}, where new composite indexes can be constructed while regressive ones are systematically removed. This coordinated workflow enables the system to leverage the strengths of both heuristics, leading to a more robust, stable, and effective index recommendation pipeline than relying on any single mechanism.

\begin{figure*}[htp]
  \centering
  \includegraphics[width=\linewidth]{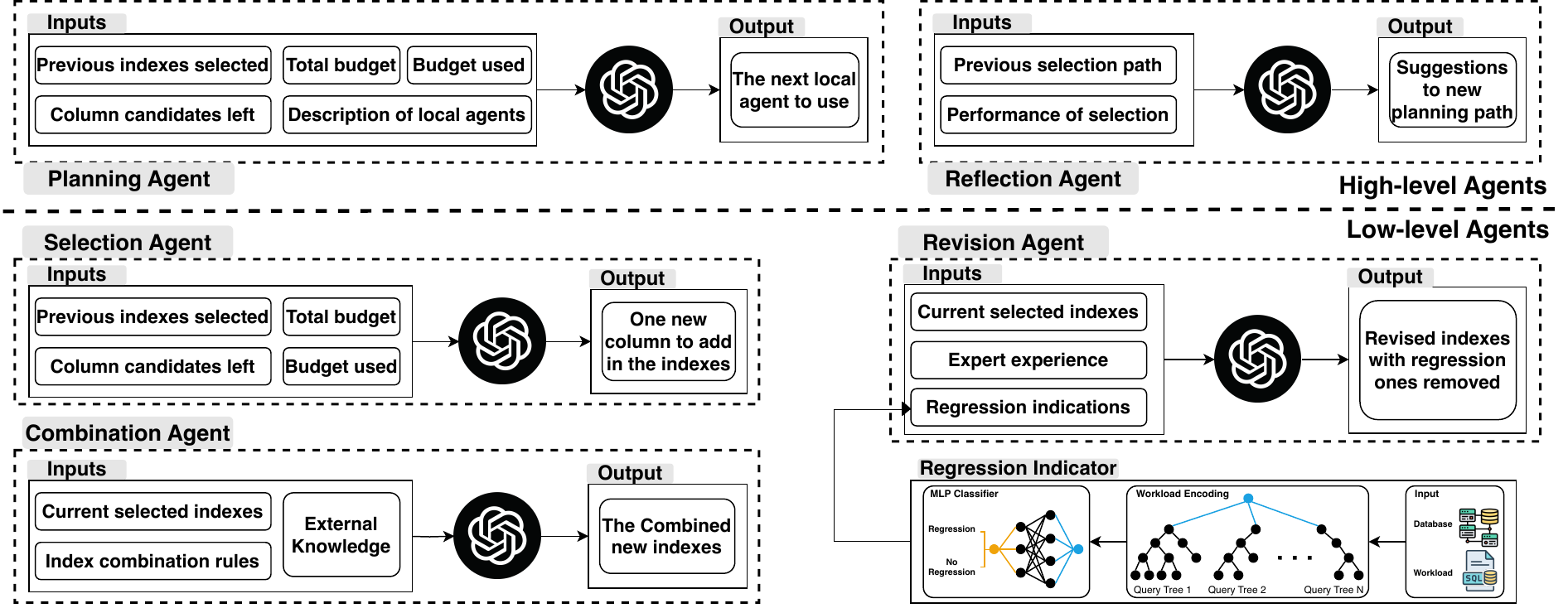}
  \caption{Overview of the inputs and outputs of our high-level and low-level agents adopted in the multi-agent pipeline.}
  \Description{agents}
  \label{fig:agents}
\end{figure*}

\section{Workload Representation}
\label{5}
In this section, we will introduce in detail our efficient workload representation. In Section~\ref{sec:51}, we will first introduce the information we extract as a representation for each query in the workload. Then in Section~\ref{sec:52}, we
will explain how we further enhance pipeline efficiency by reformulating and merging the information in the whole workload into column candidates.

\subsection{Query Information Extraction}
\label{sec:51}

To fully leverage the reasoning capabilities of LLMs, the input information should be concise and readily interpretable. However, data extracted directly from SQL queries is often highly structured and domain-specific, making it challenging for LLMs to process effectively. Therefore, it is necessary to extract and reformulate query-related information into a format that LLMs can interpret and process efficiently.
Since we need to build index candidates derived from related columns, we first extract column information from workload queries. Building on insights from prior studies~\cite{qpeval, llmidxadvis, distill} and our preliminary experiments, we identify that the essential column-related information includes:

\vspace{-\topsep}
\begin{itemize}[leftmargin=*]
\setlength{\itemsep}{0pt}
\setlength{\parsep}{0pt}
\setlength{\parskip}{0pt}
  \item \textbf{Indexable column names:} Table and column names are the key attributes for constructing indexes. We define each column that appears in the workload as an `indexable column'. Specifically, the format `TableName.ColumnName' will be used in the column candidate for LLMs to choose during the selection phase;
  \item \textbf{Column operator type:} Operators play important roles in index recommendation. Indexes can post a significant change in the efficiency of an operation. The operators related to the indexable column can be easily accessed in a query plan;
  \item \textbf{Estimated column cardinality:} As one of the key attributes affecting index selection, cardinality represents the uniqueness of data values contained in a column. Generally, indexes on higher cardinality columns allow easier access to specific rows. However, indexes on low cardinality columns may also be beneficial in specific situations, such as when the queried column value is extremely infrequent compared to others in the same column;
  \item \textbf{Column distribution:} Column distribution includes data histogram, minimum, and maximum values within the column. Combining the column distribution and column cardinality, LLM can have better discrimination of more and less effective indexes;
  \item \textbf{Estimated column storage:} To optimize the efficiency of the indexes selected, it is necessary to keep track of the storage cost of the current index recommendation. Although the real storage cost can not be calculated before the actual index creation, the database optimizers can estimate the storage using hypothetical indexes~\cite{hypopg};
  \item \textbf{Estimated column utility:} Apart from the information above, an estimation of the potential improvement from a column index is vital for telling if a column candidate is worth selecting~\cite{distill}. We define the utility of a potential index on a query as follows:
  \begin{definition}
    (Index Utility): Given an index $I$ and a query $q$, the utility $U_I(q)$ of creating the index $I$ for $q$ is defined by query total cost without and with index $I$ per unit of storage cost: 
    \vspace{-\topsep}
    \begin{displaymath}
    \small
       U_I(q) = \frac{Est.\ Cost_{w/o\  I} (q) - Est.\  Cost_{w/\  I}(q)}{Est.\  Storage(I)}
    \end{displaymath}
    where database optimisor uses $Est.\ Cost(q)$ to estimate the total time cost of the query $q$, and uses $Est.\  Storage(I)$ to estimate the storage cost of the index $I$.
  \end{definition}
  The utility of a column $c$ on query $q$ will then be calculated as $U_c(q)$. From the calculation, the estimated column utility is designed to be compound information, including column cardinality, query selectivity, and operations on the column.
\end{itemize}
\vspace{-\topsep}

\subsection{Workload Information Summarization}
\label{sec:52}
\textbf{Motivation.} Given a workload with $m$ queries, we could extract $m \times n$ column candidates by individually processing each query and extracting an average of $n$ indexable columns per query. However, this approach is inefficient because many columns appear repeatedly across different queries, creating numerous duplicates. To enable the LLM to make effective and efficient index selections, we must consolidate this information so that each candidate corresponds to a unique column rather than being repeated across multiple queries.

To achieve this goal, we aggregate information for columns with the same name, ensuring that our final candidate set contains only unique columns. For most column attributes, consolidation is straightforward. The \textit{estimated column cardinality}, \textit{column distribution}, and \textit{estimated column storage} remain constant for a given column regardless of which query references it, so we retain these values directly.
However, two attributes vary across queries and require special handling: 

\noindent (1) \textit{column operator types}: We merge these by collecting all operators observed for a column across different queries into a single comprehensive list.

\noindent (2) \textit{estimated column utility}: This requires a more principled aggregation approach. Since the goal of index recommendation is to enhance overall workload performance, columns that provide greater reductions in query execution cost should be prioritized. Therefore, we aggregate utility values across all queries that reference a given column. A higher aggregated utility indicates that indexing this column will likely yield greater overall performance improvement for the entire workload.

\section{Multi-agent Pipeline}
\label{6}

This section presents our hierarchical multi-agent pipeline for index recommendation, decomposing the problem into structured sub-steps handled by specialized agents. Section~\ref{sec:61} introduces three low-level agents that select, combine, and revise indexes within individual sub-steps. Section~\ref{sec:62} details two high-level agents responsible for planning and resource allocation under budget constraints. Section~\ref{sec:63} demonstrates how these agents interact to achieve budget-aware planning and coordination.

\subsection{Low-level Agents}
\label{sec:61}
\textbf{Selection agent}: As the key function of index advisor, the selection agent plays the role of selecting new columns from the column candidates given. As shown in Figure~\ref{fig:agents}, the inputs for the selection agent include previous selected indexes, column candidates to select from, total storage budget, and the storage budget already used. Once the selection agent is called, LLM is prompted to select one new column into the current index recommendation as a single-column index. 

\noindent\textbf{Combination Agent}: As noted earlier, the selection agent produces only single-column indexes, which are often insufficient when queries involve multiple columns from the same table. In such cases, composite indexes can provide greater efficiency than maintaining several separate single-column indexes. To address this, we introduce the combination agent, which merges selected single-column indexes into composite indexes that better capture multi-column access patterns. In doing so, it also mitigates optimizer estimation bias by transforming individually suboptimal indexes into beneficial composite ones.

\begin{figure}[t]
  \centering
  \includegraphics[width=\linewidth]{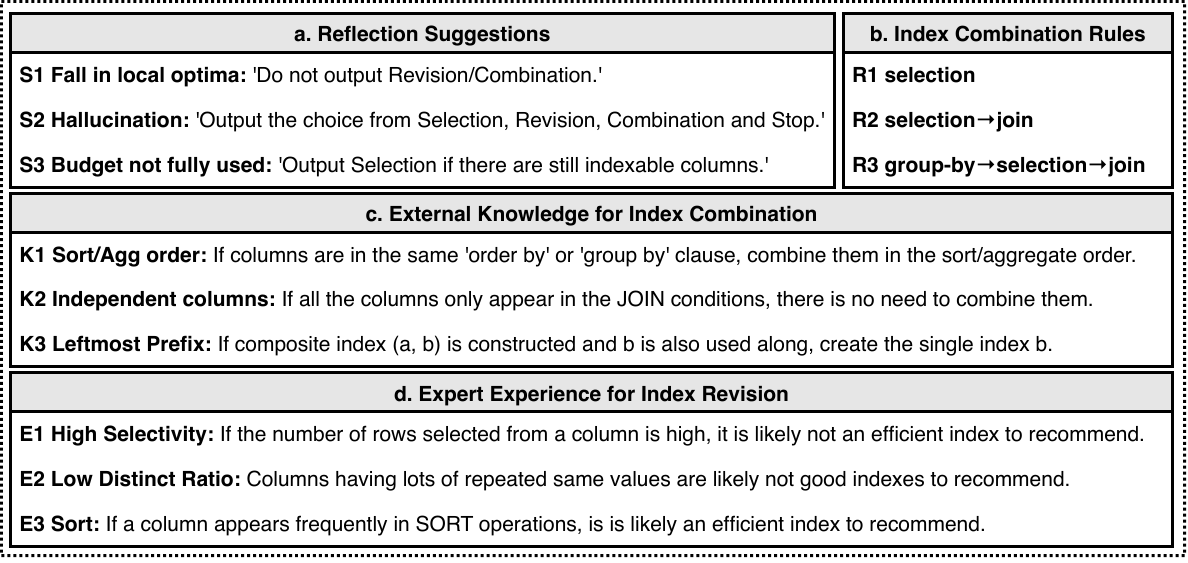}
  \caption{Examples of the agents' additional information.}
  \Description{prompt examples}
  \label{fig:prompt_examples.pdf}
\end{figure}

To guide this process, we first adopt the approach in~\cite{distill}, which outlines rules for combining indexes, including valid column groupings and orderings inferred from query structures, as illustrated in Figure~\ref{fig:prompt_examples.pdf}(b). Beyond combination rules, we further enhance the agent with external knowledge from sources such as textbooks, expert practices, and the Internet, enabling it to incorporate domain expertise into index construction. Examples of such knowledge are shown in Figure~\ref{fig:prompt_examples.pdf}(c). As shown in Figure~\ref{fig:agents}, given the current indexes, query operations, combination rules, and external knowledge, the combination agent determines whether certain indexes should be merged, and outputs updated recommendations.


\begin{figure}[t]
  \centering
  \includegraphics[width=\linewidth]{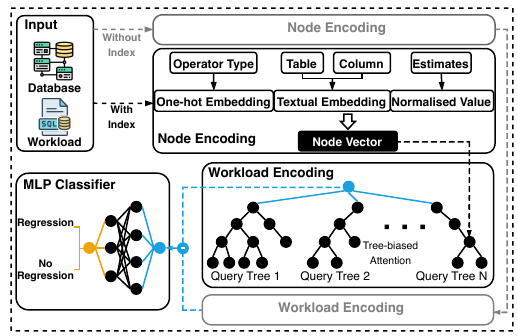}
  \caption{The structure of the regression indicator. Potential index regression will be predicted using the workload representation difference before and after the indexes.}
  \Description{regression_indicator}
  \label{fig:regression_indicator}
\end{figure}

\noindent\textbf{Revision Agent}: 
The revision agent will be adopted to eliminate the hallucinations and potentially inefficient indexes in the current index recommendation, which may lead to index regression. First, a hallucination check will be conducted in the revision agent to identify and correct the wrong or non-existing index selections. Second, to mitigate the impact of potentially inaccurate estimations made by the database optimizer, a problem that is well documented even in traditional optimization pipelines, the LLM is guided by a three-fold justification mechanism that draws upon \textbf{column information}, \textbf{expert experience} and \textbf{regression indications}, as shown in Figure~\ref{fig:agents}. Apart from the optimizer estimations, LLMs can also obtain \textbf{cardinality-related column information} from the name and distribution of the candidate columns selected. For example, the cardinality for the column `country\_name' is highly likely to be smaller than the column `order\_id'. Therefore, we prompt the LLM to identify and remove columns from the selected index when a significant discrepancy exists between the estimated cardinality and the LLM’s analysis. To enhance LLM's ability to tell a regressing index, we inject \textbf{expert knowledge} drawn from diverse sources such as common practices and online discussions. Insights of inefficient indexes are then distilled into expert experience rules, which guide the detection of regressive indexes. An example of such rules is shown in Figure~\ref{fig:prompt_examples.pdf}(d).


To strengthen the LLM’s capacity for independent and robust reasoning beyond reliance on the expert experiences, we also adopt a regression indicator to provide the LLM with an objective, data-driven measure that complements LLM reasoning.

\noindent\textbf{Regression indicator}: Since prior learned index regression indicators~\cite{aimeetsai, jia} are dataset-specific and only detect query-level regressions, we propose a workload-level regression indicator which is trained using information invariant to databases and workloads. The structure of the model is presented in Figure~\ref{fig:regression_indicator}. The workload’s query plans, both with and without the indexes, are extracted as inputs for the indicator. Each operator in a plan is represented as a node, forming a binary tree for the entire query. Following the approach in~\cite{queryformer}, we encode each query tree using tree-biased attention, where a node attends only to those within its sub-tree. The root node then serves as the vector representation of the entire query. The workload representation is constructed by concatenating the root vectors of all queries. Finally, given the pair of workload vectors before and after index creation, we employ an MLP classifier to predict potential index regressions based on the difference between the two representations. Each index will be assigned a regressive score predicted that the LLM can use as a reference.

Inspired by~\cite{qpeval}, we extract only the essential features from each tree node that influence index performance, including the operator type, table, column, estimated cardinality, and cost values. Since operators and estimations are inherently database-invariant, they are encoded using one-hot representations and normalization. To enhance the robustness of the regression indicator, we employ Sentence Transformers~\cite{reimers-2019-sentence-bert} to generate database-invariant embeddings for the table and column information.

\subsection{High-level Agents}
\label{sec:62}
\textbf{Planning agent:} The planning agent plays an important role in the multi-agent pipeline for managing problem sub-steps and determining which Low-level agents to deploy at each step. As shown in Figure~\ref{fig:agents}, given the column candidates, current indexes, total storage budget, and budget already used, the planning agent first evaluates the stop criteria. If these criteria are not met, it outputs an action for the current sub-step: `Selection', `Combination', or `Revision', each invoking a corresponding low-level agent. When the planning agent determines that the index selection is optimized or that no additional columns can be chosen within the budget, it triggers the `Stop' action to conclude the optimization.

\noindent\textbf{Reflection agent:} In practical use, it has been observed that planning agent's strategies for low-level agents can be inefficient at times, as it may suffer from hallucinations or become trapped in local optima. For instance, the planning agent may repeatedly revise the current index recommendation, invoke the combination agent when no viable combinations exist, or produce invalid actions due to hallucinations. To enhance the planning process, a reflection agent is introduced to evaluate the current path, with the goal of identifying and correcting errors. This helps the planning agent escape local optima and supports more efficient sub-step planning. Given the sequence of previous sub-step actions and the utility of current indexes, the reflection agent will offer suggestions for the next step's planning. Example of possible suggestions can also be found in Figure~\ref{fig:prompt_examples.pdf}(a).

\begin{figure}[htp]
\begin{algorithm}[H]
\caption{Multi-agent Coordination}\label{alg:ma-pipeline}
 \small{
\begin{algorithmic}[1]
\Require Column candidates $c_i = \{N_i, T_i, E_i, D_i, S_i, U_i\}\in C$
\Require Total storage budget $B$
\Require Planning agent $Plan()$, Reflection agent $Reflect()$
\Require Selection agent $Select()$,  Combination agent $Combine()$,  Revision agent $Revise()$
\Require Maximum number of sub-steps $\alpha$
\State $a = 0$ \Comment{Initial agent call}
\State $B_a = 0$ \Comment{Initial storage budget used}
\State $I_a = []$ \Comment{Initialize index selection}
\State $Actions = []$ \Comment{Initialize planning path}
\State $p_a = None$ \Comment{Initialize planning agent output} 
\While{$a \leq \alpha$ and $p_a \neq \text{'Stop'}$}      
\State $p_a = Plan(B, C, I_a, B_a, Suggestion)$  \Comment{Call the planning agent}
\If{$p_a == \text{'Selection'}$} \Comment{Adopt selection agent}
    \State $c_i = Select(B, C, I_a, B_a)$ \Comment{Select a new single-column index}
    \State $Append(I_a, c_i)$  \Comment{Append to selected indexes}
    \State $B_a = B_a + S_i$ \Comment{Update storage budget used}
\ElsIf{$p_a == \text{'Combination'}$}
    \State $I_a = Combine(I_a)$ \Comment{Construct composite indexes}
    \State $B_a = Storage(I_a)$ \Comment{Update the selected indexes' storage}
\ElsIf{$p_a == \text{'Revision'}$}
    \State $I_a = Revise(I_a)$ \Comment{Revise the index recommendation}
    \State $B_a = Storage(I_a)$ \Comment{Update the selected indexes' storage}
\Else
    \State $Exception(p_a)$
    \Comment{Record an exception for undefined agent call}
\EndIf
\State $Append(Actions, p_a)$
\State $Suggestion = Reflect(Actions)$ \Comment{Get planning reflections}
\State $a = a + 1$
\EndWhile
\State Output the final index selection $I_a$
\end{algorithmic}
}
\end{algorithm}
\end{figure}

\subsection{Multi-agent Planning and Coordination}
\label{sec:63}
Once the high-level and low-level agents are developed, they can be integrated into a hierarchical multi-agent pipeline enabling efficient and effective zero-shot index recommendation. As shown in Algorithm~\ref{alg:ma-pipeline}, the multi-agent pipeline receives the storage budget and the column candidates as inputs. To balance efficiency with effectiveness, we introduce a hyperparameter $\alpha$, which defines the maximum number of planning sub-steps. At the beginning of each sub-step, the planning agent determines the next action in line 7. If additional indexes can be selected, the selection agent chooses a single-column index from the effectively represented column candidates from line 9 to line 11. When combining existing indexes into a composite index could further improve workload performance, the combination agent in line 13 refines the current recommendation by composing these indexes. Since the selected indexes may contain duplicates, regressions, or low-impact choices, the revision agent ensures effectiveness by removing such cases once called in line 16. After each action, the reflection agent reviews prior decisions and provides guidance for subsequent planning. Through the planning agent’s reasoning in line 22, if the index configuration is deemed optimal, iterations terminate early, and the final selection is output as the index recommendation.


\begin{table*}[htp]
\caption{The benefit-to-cost ratio of all methods on four different benchmarks under 2\%, 10\%, 20\% and 50\% storage budgets. Due to the DROP heuristic, the starred$^*$ methods recommend no index on some small storage budgets.}
\label{tab:main_results}
\resizebox{\linewidth}{!}{
\begin{tabular}{ccccccccccccccccc}
\toprule
Dataset   & \multicolumn{4}{c}{TPC-H}       & \multicolumn{4}{c}{TPC-DS} & \multicolumn{4}{c}{DSB} & \multicolumn{4}{c}{JOB}   \\ \cmidrule(lr){2-5}\cmidrule(lr){6-9}\cmidrule(lr){10-13}\cmidrule(lr){14-17}
Storage Budget   & 2\%  & 10\% & 20\%  & 50\%  & 2\%  & 10\% & 20\%  & 50\%  & 2\%  & 10\% & 20\%  & 50\%  & 2\%  & 10\% & 20\%  & 50\%\\ \hline \hline
Anytime  & 0.089          & -0.010         & 0.415          & 0.198          & \textbf{15.285} & 4.588          & 1.494          & 0.821 & 0.063          & 0.040          & 0.019          & 0.147          & 0.979          & -0.297         & 0.095          & 0.080           \\ 
AutoAdmin$^*$  & -  & 0.714  & 0.714  & 0.202 & 0.746  & 0.768  & 0.080 & 0.047 & 0.130 & 0.022 & 0.017 & 0.006 & - & -  & 0.129 & 0.066 \\ 
DB2Advis   & 0.007          & -0.034         & -0.012         & 0.144          & 7.342           & 2.961          & 1.364          & 0.784 & 0.092          & 0.054          & 0.573          & 0.167          & 1.007          & 0.227          & \textbf{0.162} & 0.060    \\ 
Drop$^*$   & -          & \textbf{0.716} & \textbf{0.715} & 0.154          & 0.749           & 0.767          & 0.102          & 0.117 & 0.201          & 0.021          & 0.008          & 0.103          & -          & -          & 0.132          & 0.088 \\ 
Extend & -0.013         & -0.029         & 0.466          & 0.226          & 2.136           & 2.257          & 0.864          & 0.702 & 0.061          & 0.044          & 0.013          & 0.169          & 1.111          & 0.245          & 0.099          & 0.067  \\ 
Relaxation    & 0.080          & 0.007          & 0.444          & 0.218          & 10.141          & 2.561          & \textbf{1.969} & 0.709 & 0.068          & 0.041          & 0.012          & 0.149          & \textbf{1.378} & 0.240          & 0.086          & 0.071 \\ \hline
SWIRL  & 0.010          & 0.027          & 0.496          & 0.172          & 0.629          & -0.557           & 0.496          & 0.153          & 0.059 & 0.022 & 0.211 & 0.007   & 0.279          & 0.097 & 0.113 & 0.080   \\
BALANCE    & 0.023 & 0.045 & 0.496 & 0.158 & 4.035 & 0.097   & 0.732 & 0.222 & 0.043 & 0.041          & 0.207          & 0.170   & 0.273 & 0.114 & 0.110          & 0.077    \\ \hline
LLMIdxAdvis   & -0.013         & -0.025         & 0.684          & 0.203          & 2.136          & 0.754          & 1.642          & 0.394        & 0.061          & -0.630         & 0.246          & 0.250            & 0.943          & 0.202          & 0.126          & 0.090   \\ \hline
AMAZe  & \textbf{0.220} & 0.587          & 0.587          & \textbf{0.299} & 9.726           & \textbf{4.756} & 1.594    &  \textbf{1.594}     & \textbf{0.919} & \textbf{1.224} & \textbf{0.598} & \textbf{0.344} & 0.617          & \textbf{0.452} & 0.125          & \textbf{0.094} \\ 
\bottomrule
\end{tabular}}
\end{table*}

\section{Experiment}
In this section, we will evaluate our multi-agent index advisor's performance in terms of effectiveness, efficiency, and zero-shot generalisation ability. 

\subsection{Experimental Setup} 
\label{sec:71}
\textbf{Dataset.} We select four different datasets from both hypothetical and real-world settings for our evaluations.

\textit{\underline{TPC-H:}} We generate the TPC-H data with a scale of 10 (around 10 GB). The TPC-H benchmark has in total 22 different query templates. In order to simplify the analysis and make a fair comparison of the methods, we follow previous methods~\cite{distill, extend} to remove the query templates 2, 17, 20 that contribute significantly to the total execution time of the workload or have little potential improvement from indexes. The test workload for TPC-H has in total of 19 queries randomly generated from the 19 query templates left.

\textit{\underline{TPC-DS:}} TPC-DS data with a scale of 10 (around 10 GB) is generated. The TPC-DS benchmark has 99 different query templates, and for the same reason in the TPC-H benchmark, we remove the query templates 4, 6, 9, 10, 11, 14, 23, 24, 32, 35, 36, 39, 41, 70, 86, and 95. The test workload for TPC-DS consists of 83 queries generated from the 83 query templates left.

\textit{\underline{DSB:}} The DSB benchmark~\cite{ding2021dsb} is derived from the TPC-DS benchmark, while enhanced with complex data distribution and challenging query templates. DSB has in total 15 templates for single block queries, 22 templates for multi-block queries, and another 15 templates for SPJ queries, especially evaluating techniques with limited capabilities. In our experiments, we only utilized the general-purpose 37 query templates in the first two groups and constructed a workload of 37 queries randomly generated from the templates.

\textit{\underline{IMDB (JOB workload):}} Join order benchmark (JOB~\cite{Leis2015HowGA}) is an OLAP benchmark querying the IMDB~\cite{maas-EtAl:2011:ACL-HLT2011} dataset and designed to assess the capabilities of a database management system (DBMS) in optimizing and executing complex real-world join queries. The JOB workload has 113 real-world queries.

\noindent \textbf{Experiment Environments.} We choose the popular DBMS PostgreSQL 14.2 to evaluate all the queries~\cite{PostgreSQL}. We follow the official documentaries and websites~\cite{TPC-H.tools, Leis2015HowGA, dsbgit} to create PostgreSQL databases for all the above datasets. All the experiments are conducted on an isolated server with two Intel(R) Xeon(R) Platinum 8469C CPUs featuring 46 cores and 96 threads, along with 460GB of RAM.

\noindent \textbf{Baseline Methods.}
\textit{\underline{(1) Heuristic-based baselines}}: Heuristic-based index advisors can be broadly categorized into two types by their optimization approaches. Those following an ADD strategy incrementally build a set of indexes by selecting candidates one by one. In our experiments, we include Anytime~\cite{chaudhuri2020anytime}, DB2Advis~\cite{10.5555/846219.847390}, Extend~\cite{extend}, and Relaxation~\cite{10.1145/1066157.1066184} as representatives of this group. On the other hand, AutoAdmin~\cite{10.5555/645923.673646} and Drop~\cite{315323} serve as baselines for the DROP strategy, where advisors start with an initial set of indexes and iteratively eliminate those deemed inefficient. \lzdh{We reproduce all the heuristic-based baselines based on the evaluation work~\cite{magicmirror}.} \textit{\underline{(2) Learning-based baselines}}: We adopt two RL-based index advisors as our baselines, namely SWIRL~\cite{swirl} and BALANCE~\cite{balance}. While SWIRL includes vast amounts of workloads during training to largely shorten the inference time, BALANCE utilizes transfer RL to increase the generalisability to the dynamic workload scenarios. \textit{\underline{(3) Prompt-based baselines}}: The recent work LLMIdxAdvis~\cite{llmidxadvis} proposes to recommend indexes by prompting the LLM with prepared demonstrations. Selecting the best demonstration for the input workload and calling the LLM helps LLMIdxAdvis achieve higher efficiency and generalisability. Due to the high computation cost for preparing new demonstrations for our workloads, we directly use LLMIdxAdvis' prepared demonstrations since they cover all the databases in our experiment.

\noindent \textbf{Evaluation metrics.} We adopt two metrics for our evaluations. The \textbf{benefit-to-cost ratio} metric is first introduced by~\cite{10.5555/846219.847390} and evaluates the absolute workload execution time reduction per unit storage used. The larger the benefit-to-cost ratio, the better an index advisor performs in both workload cost reduction and efficient storage use. The \textbf{relative workload improvement} is defined as the percentage of the workload's actual execution time decrease before and after the construction of the recommended indexes. We measure all the workload execution time in seconds and storage cost in MegaBytes (MB).

\noindent \textbf{LLM and Training Setting.} We leverage the GPT-4o-0806~\cite{openai2023gpt4} version within the OpenAI API as the default LLM backbone for the agents. Furthermore, we also assess our system’s generalizability across other LLMs, as detailed in Section~\ref{sec:83}. For the index indicator, we train the model on a server with one Nvidia A100 GPU.

\begin{figure*}[htp]
  \centering  \includegraphics[width=\linewidth]{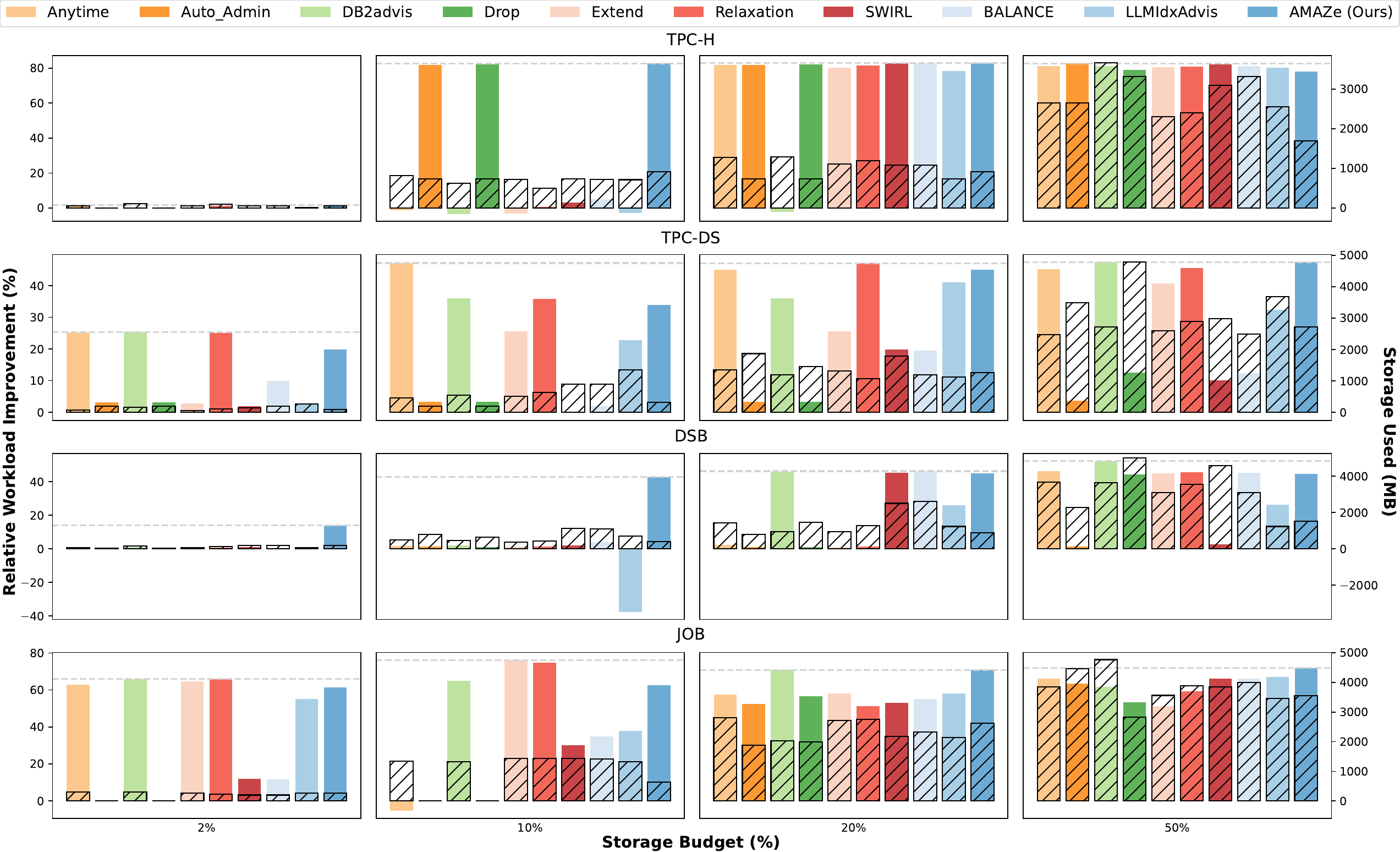}
  \caption{Method relative workload improvement (colored) v.s. storage used (shaded) under different storage budgets.}
  \Description{results_figure}
  \label{fig:results_figure}
\end{figure*}

\subsection{Generalization and Effectiveness Evaluation}
\label{sec:72}
As shown in Table~\ref{tab:main_results}, we compare our proposed AMAZe's benefit-to-cost ratio with the baselines on four benchmarks. All the methods are evaluated with four different storage budgets, taking 2\%, 10\%, 20\%, and 50\% of the database size. Our results reveal several key observations:

\noindent\textbf{(1)} \textbf{AMAZe outperforms all the baselines in most of the situations.} \lzdh{The benefit-to-cost ratio for our method is the highest in 10 out of the 16 experimental settings, and remains the top tier for the rest 6 settings, especially compared to the heuristic-based ßand prompt-based methods, which are also zero-shot}. Specifically, AMAZe doubles the ratios from the SOTA performance given 2\%, 10\%, and 50\% storage budgets in the most complicated DSB benchmark. Similar improvement can be observed in the 10\% and 50\% budget settings for the real-world benchmark JOB. This demonstrates \textbf{AMAZe's strong generalization capability to a variety of database schemas and workloads in zero-shot settings}.

\noindent\textbf{(2)} \textbf{AMAZe successfully alleviates the index regression problem}, 
which commonly occurs in the zero-shot heuristic-based methods. For example, Anytime suffers from the index regression in TPC-H and DSB under 10\% budget, and similar problems can be observed for Extend in TPC-H and for SWIRL in TPC-DS under 10\% budget. Our detailed analysis reveals that  heuristic-based methods highly rely on the benefit estimations from optimizers, index regressions always happen when the index advisors have a limited storage budget and select the regressive indexes. For example, DB2Advis recommends indexes for TPC-H that lead to a nearly 50\% execution time regression on query template 9 from 14.76 seconds to 26.60 seconds. Among the 13 indexable columns in query 9, only two columns appear in DB2Advis' index recommendation. In addition, the composite indexes DB2Advis selects `I(C supplier.s\_nationkey,C supplier.s\_suppkey)' and `I(C part.p\_type,C part.p\_partkey)' are not efficient for JOIN conditions `ps\_suppkey = l\_suppkey' and `p\_partkey = l\_partkey'. Unfortunately, the improvements for the indexes on other queries are not significant and therefore result in an overall regression. In contrast, our AMAZe mitigates the problem with the help of the specifically designed workload representation technique and low-level revision agent.

\noindent\textbf{(3)} \textbf{AMAZe mitigates the training instability or hallucination problem faced by learning-based / prompt-based methods}. Our results show that the performance for the learning-based index advisors suffers from higher fluctuation across benchmarks. As an example, while BALANCE performs well on TPC-DS and DSB, it fails to beat heuristic-based baselines in the rest of the benchmarks due to the highly variable training process~\cite{breakitdown}. Similar variation happens in the prompt-based methods due to the hallucination of LLMs. As an example, LLMIdxAdvis' performance on TPC-DS 20\% budget is among the top tier, but the performance on the other budget settings is low. In contrast, our hierarchical multi-agent pipeline verifies and corrects the potential hallucinations during the index recommendation, effectively mitigating the problem.


In order to further analyze the capability of our method, we plot the two metrics we used to compute the benefit-to-cost ratio, namely the relative workload improvement and the storage cost. From Figure~\ref{fig:results_figure}, we can observe that:

\noindent\textbf{(1)} \textbf{AMAZe achieves the highest relative workload improvement in 8 out of in total 16 experiment settings.} Given smaller storage budgets from 2\% to 20\%, AMAZe tends to recommend indexes that lead to the most workload relative reduction. However, for a larger budget like 50\%. Our method prefers indexes with comparable improvement but much lower storage cost. It is also observed that in the experiment settings where our method did not achieve the highest benefit-to-cost ratio, for example, TPC-H 20\% budget and JOB 20\% budget, AMAZe can still recommend indexes that result in the largest workload execution time decrease, with a trade-off of higher storage cost.

\noindent\textbf{(2)} \textbf{AMAZe achieves SOTA performance even for small storage budgets.} In contrast, heuristic-based baselines following the DROP heuristic fail to perform well for small storage budgets. The main reason is that the small-storage indexes are generally not as beneficial as large-storage indexes and are removed from the recommendation in early DROP stages. Our AMAZe assembles a hybrid of both ADD and DROP heuristics that iteratively selects and revises the index recommendation by multi-agent coordinations.

\noindent\textbf{(3)} \textbf{AMAZe with the multi-agent design outperforms the prompt-based baseline, especially in the most complicated DSB.} The prompt-based LLMIdxAdvis performs well on TPC-H and JOB benchmarks but has a lower performance on complicated workloads and database schemas, like TPC-DS and DSB. AMAZe's better performance comes from the efficient workload representation technique which summarizes the information from workloads and database schemas. In addition, the multi-agent pipeline decomposes the problem and simplifies the task for each LLM agent.  

\noindent\textbf{(4)} \textbf{AMAZe better controls the storage budget usage during the  recommendation.} It is observed that LLMIdxAdvis struggles to control the storage budget used in TPC-DS and DSB. As an example in TPC-DS, LLMIdxAdvis uses 1356.38 MB under 10\% budget, which exceeds the budget, and a lower 1122.97 MB under 20\% budget. The main reason is that although the available storage budget is an input in LLMIdxAdvis, LLMs are not given the storage cost when selecting new indexes. In contrast, AMAZe's planning and selection of indexes is always storage-aware since the estimated storage cost is recorded in each one of the column candidates.

\begin{table*}[htp]
\caption{Method inference runtime, the starred$^*$ learning-based methods actually require much longer time to train before applying to a benchmark.}
\label{tab:latency}
\resizebox{\linewidth}{!}{
\begin{tabular}{ccccccccccccccccc}
\toprule
Dataset   & \multicolumn{4}{c}{TPC-H}       & \multicolumn{4}{c}{TPC-DS} & \multicolumn{4}{c}{DSB} & \multicolumn{4}{c}{JOB}   \\ \cmidrule(lr){2-5}\cmidrule(lr){6-9}\cmidrule(lr){10-13}\cmidrule(lr){14-17}
Storage Budget   & 2\%  & 10\% & 20\%  & 50\%  & 2\%  & 10\% & 20\%  & 50\%  & 2\%  & 10\% & 20\%  & 50\%  & 2\%  & 10\% & 20\%  & 50\%\\ \hline \hline
Anytime    & 17.07         & 24.85          & 42.26         & 84.49 & 1846.86       & 1853.05          & 1850.87       & 1861.07       & 1741.27       & 1874.02       & 1865.24       & 1876.40          & 1850.46       & 1837.55        & 2000.60       & 2618.21    \\ 
AutoAdmin  & -     & \textbf{0.47}  & 3.41          & 20.01          & 12.42         & 50.47            & 50.47         & 168.11        & 10.05         & 73.84         & 73.84         & 135.90           & -             & -              & 61.93         & 1839.31     \\ 
DB2Advis    & 14.31         & 23.81          & 23.36         & 23.49          & 46.00         & 44.63            & 45.58         & 47.71         & 46.81         & 48.21         & 50.22         & 47.80            & 40.58         & 33.72 & 33.88         & 36.54       \\ 
Drop   & -      & 12.00 & 26.57         & 80.71          & 2243.51       & 2241.48          & 2241.48       & 2229.48       & 2026.49       & 2229.92       & 2229.92       & 2229.48          & -             & -              & 1869.58       & 1849.99     \\ 
Extend        & 16.43         & 17.12          & 14.65         & 23.48          & 57.58         & 119.71           & 192.50        & 359.81        & 55.92         & 96.41         & 136.26        & 309.31           & 91.68         & 362.16         & 693.30        & 1520.86     \\ 
Relaxation      & 40.66         & 39.44          & 43.39         & 36.10          & 4987.62       & 4917.65 & 4536.28       & 4356.80       & 1469.11       & 1476.62       & 1385.13       & 1285.31 & 6952.19       & 6708.86        & 6538.98       & 5846.26  \\ \hline
SWIRL$^*$     & 1.09          & 1.30           & 2.44          & 2.42           & 3.73          & 3.35             & 3.68          & 3.71          & 2.59 & \textbf{1.88} & \textbf{2.78} & \textbf{2.06}    & 4.69          & 2.95  & \textbf{4.56} & \textbf{4.26}  \\
BALANCE$^*$   & \textbf{1.03} & 1.36  & \textbf{2.34} & \textbf{2.28}  & \textbf{3.40} & \textbf{3.31}    & \textbf{3.61} & \textbf{3.44} & \textbf{2.47} & 1.96          & 2.86          & 2.15    & \textbf{4.45} & \textbf{2.86}  & 4.77          & 4.31     \\ \hline
LLMIdxAdvis    & 210.23        & 148.95         & 216.63        & 203.83         & 153.35        & 167.76           & 412.48        & 138.95        & 134.77        & 123.45        & 146.10        & 177.62           & 141.86        & 165.49         & 170.04        & 174.96      \\ \hline
AMAZe & 31.85         & 44.31          & 122.21        & 156.82         & 51.00         & 85.75            & 106.71        & 133.21        & 43.23         & 83.67         & 115.38        & 178.82           & 69.22         & 66.18          & 106.62        & 211.56  \\ 
\bottomrule
\end{tabular}}
\end{table*}

\subsection{Efficiency Evaluation}
\label{sec:73}
Table~\ref{tab:latency} reports the algorithm inference runtime for all the methods to evaluate their efficiency. Through analyzing the inference runtime of different methods, we make  the following observations:

\noindent\textbf{(1)} \textbf{AMAZe's 
runtime outperforms the prompt-based baseline LLMIdxAdvis in most of the settings}. The reason is that with the multi-agent pipeline, our method’s main computational overhead comes from invoking the LLM API and updating column candidates, whereas LLMIdxAdvis faces an efficiency bottleneck due to the added cost of identifying the best demonstrations for each selection. In addition, LLMIdxAdvis suffers from high latency during its demonstration construction stage.  17 hours and \$90 are reported to be taken to collect the demonstrations~\cite{llmidxadvis}. It is also note-worthy that LLMIdxAdvis' time cost, given increasing storage budgets, does not change a lot since they call the same demonstration match algorithm. In contrast, 
AMAZe's 
runtime may increase as the budget increases. It is because our index recommendation is storage-aware. The more budget given, the more low-level agents are planned by the planning agent. However, AMAZe's time cost under the largest 50\% budget is still lower than LLMIdxAdvis.

\noindent\textbf{(2)} The learning-based methods have significantly faster inference time compared to the other methods. The reason is that the learning-based methods only need to call the trained models with low latency during inference. However, in a zero-shot situation, the training time of the learned models shall also be considered as part of the algorithm latency. As an example, SWIRL's training times for the TPC-H, TPC-DS, and JOB datasets are 1226.136, 3373.32, and 3635.00 seconds, respectively. In addition, both learned methods require an initial training set of data collected from the database, which is difficult to acquire in a zero-shot setting.

\noindent\textbf{(3)} \textbf{AMAZe presents a significant efficiency advantage for larger workloads and more complicated database schemas.} Heuristic-based advisors on larger workloads generally suffer from much higher latency. As an example, the algorithm runtime for Relaxation on TPC-DS with 83 queries' workload is more than 100 times the runtime on the TPC-H benchmark with 19 queries. Although the DB2Advis baseline shows a shorter runtime compared to our AMAZe, it trades off with a lower workload performance by using a simpler greedy optimization.

\noindent\textbf{(4)} It is important to note that for the methods AutoAdmin and Drop following the DROP heuristic, the hyperparameter constraint for these methods is designed to be the maximum number of indexes to recommend. While under a storage budget constraint, the number of indexes to recommend is unknown; DBAs usually begin with a larger index number constraint and then locate the best one under the storage constraint. Therefore, the algorithm time we record, which is only the time for recommending the right index, is much shorter than the actual time required to obtain the final index recommendation from scratch. As an example, the total time of recommending indexes for TPC-DS under a 10\% budget is 689.05 seconds for AutoAdmin and 5334.80 seconds for Drop.

We further evaluate the performance of AMAZe with the LLM baseline LLMIdxAdvis in terms of monetary cost. The average number of input tokens and the corresponding API cost for AMAZe and LLMIdxAdvis are recorded in Table~\ref{tab:cost}. It is observed that LLMIdxAdvis' input lengths are more stable across benchmarks. The main reason is that LLMIdxAdvis uses a similar number of demonstrations in all benchmarks. In contrast, AMAZe's input length depends on database complication and the number of column candidates. As an example, AMAZe takes 4 times more tokens for TPC-DS with 87 column candidates than TPC-H with 15 column candidates. Considering the complexity of index recommendation task, AMAZe's higher cost on more complicated workloads is 
acceptable.

\begin{table}[t]
\caption{Comparison of average number of input tokens and API cost in USD for AMAZe and LLMIdxAdvis.}
\label{tab:cost}
\resizebox{\linewidth}{!}{
\begin{tabular}
{@{\hskip3pt}l|@{\hskip3pt}c|@{\hskip3pt}c|@{\hskip3pt}c|@{\hskip3pt}c}
\hline
\textbf{Avg. Tokens (Cost \$)}        & TPC-H & TPC-DS & DSB & JOB    \\ \hline \hline
\textbf{AMAZe}  & 12165(0.03) & 68035(0.17) &66926(0.17) & 33117(0.08)  \\ \hline
\textbf{LLMIdxAdvis}  & 38736(0.10) & 38859(0.10) & 37361(0.10) & 35619(0.09) \\ \hline
\end{tabular}}
\end{table}


\begin{figure}[t]
  \centering
  \includegraphics[width=\linewidth]{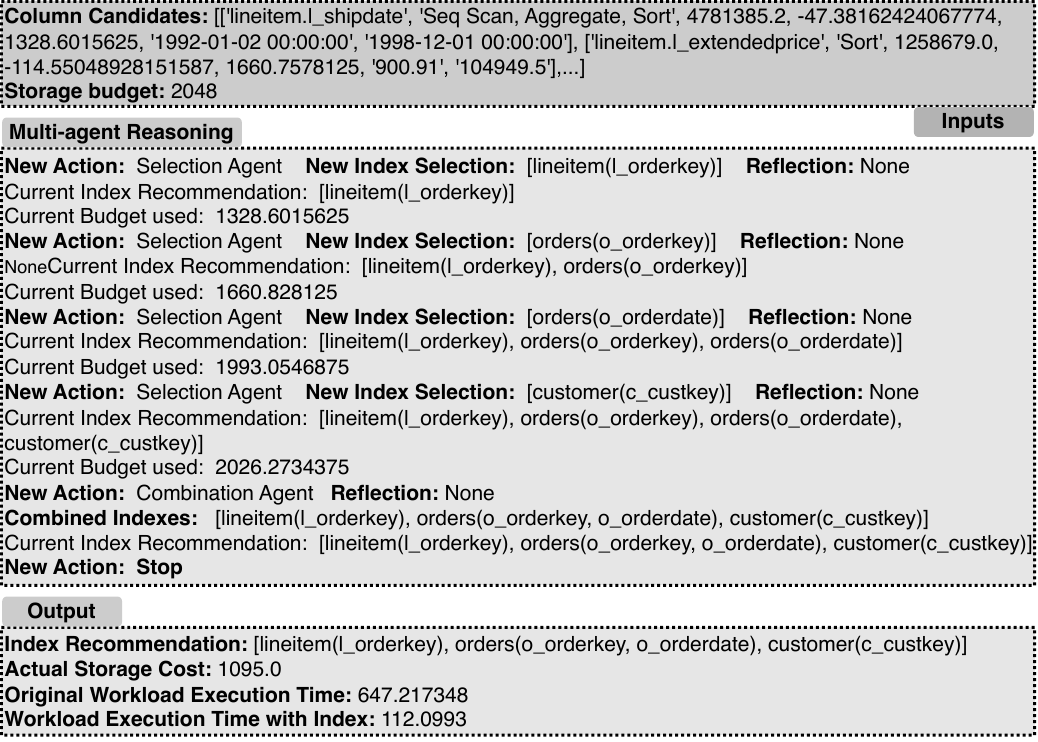}
  \caption{Qualitative example of the AMAZe's pipeline on TPC-H benchmark under 20\% storage budget.}
  \Description{pipeline_example}
  \label{fig:pipeline_example}
\end{figure}

\begin{figure*}[htp]
  \centering  \includegraphics[width=\linewidth]{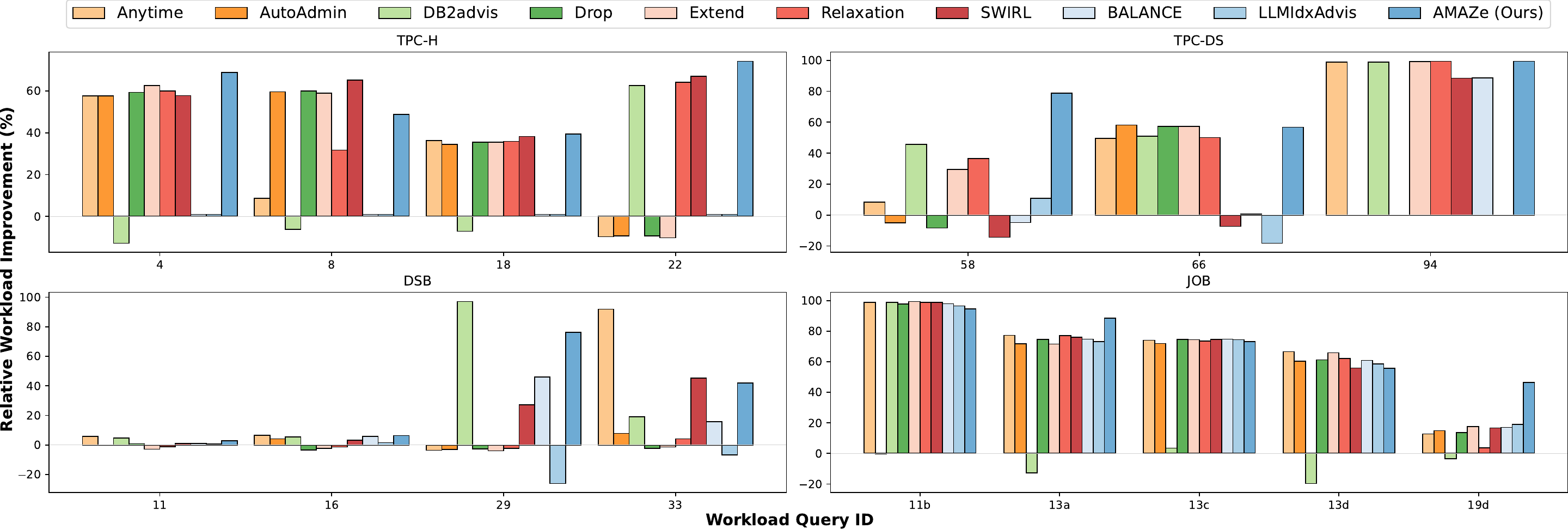}
  \caption{Relative workload improvement of the query IDs suffering from index regressions by baseline index recommendations.}
  \Description{queryid}
  \label{fig:queryid}
\end{figure*}

\subsection{Qualitative Evaluation}
\label{sec:74}

An example of AMAZe's planning steps and index recommendation on TPC-H 20\% budget can be found in Figure~\ref{fig:pipeline_example}. We observe that AMAZe efficiently recommends the indexes within five sub-steps, including selection and combination. Since the selected index is within the storage budget, the revision agent is not called by the planning agent. 
Note that AMAZe adopts the same storage cost estimation as the other baselines. Therefore, the actual storage cost of the output is smaller than the estimation in the planning phase. AMAZe can also opt to construct and calculate the actual index storage cost at each sub-step at a cost of a higher algorithm runtime. 

In order to further analyze how well AMAZe addresses the index regression problem, we pick out the queries in each benchmark having any methods' recommended indexes under 20\% budget leading to a more than one second's regression. As shown in Figure~\ref{fig:queryid}, it is observed that DB2Advis has the most regression cases, trading off the high algorithm efficiency. Most heuristic-based advisors suffer from regressions on the most complicated DSB benchmark, since their optimization relies solely on the optimizer estimations. It is important to note that, at the workload level, significant improvements on some queries with a cost of regression on some other queries is acceptable as long as the overall execution time is decreased. On the other hand, although AMAZe does not result in the highest query time improvement on all mentioned queries, the revision agent manages to avoid significant regressions and therefore leads to a higher workload-level performance.

\subsection{Ablation Studies}
In this section, we first conduct an ablation study to evaluate the effectiveness of AMAZe's workload representation technique and multi-agent pipeline. We will then analyze AMAZe's generalisation ability over the LLM backbone and workload scale. 

\noindent \textbf{Effectiveness of Workload Representation.}
To validate the effectiveness of AMAZe's workload representation step, we perform an ablation study in which the database schema and raw workload queries are directly input to the LLM agents. Instead of selecting from column candidates, the selection agent is now prompted to select one possible column index from the SQL queries. The result for this raw workload setting is presented in Table~\ref{tab:ablation}. We can observe that the performance drops significantly from 61.33\% to 29.52\%, which is reasonable since the planning and selection agents will plan and select indexes solely based on their domain knowledge and even without any demonstrations. In addition, the input context for the agents will be much longer, and LLMs' hallucination and misunderstanding will increase significantly.

\noindent \textbf{Effectiveness of Multi-agent pipeline.}
To explore the effectiveness of our multi-agent pipeline, we prompt the LLM to directly select from the column candidates and construct indexes in one step. Given the column candidates and storage budget, LLM is called only once to output all the recommended indexes. We present the one-call recommendation result in Table~\ref{tab:ablation}. 

We can observe that the one-call recommendation yields an even lower performance of 11.20\%. LLM fails to select effective indexes and control the storage budget at the same time. LLM tends to select fewer indexes with little improvement, but also little storage cost. In addition, we design another ablation with both one-call recommendation and raw workload input. The relative workload improvement further drops to 10.58\%, showing that both our workload representation technique and the multi-agent pipeline are important in solving the index recommendation problem.

\noindent \textbf{Effectiveness of Regression Indicator.}
\label{sec:83}
We further test the effectiveness of the specially trained regression indicator embedded in the revision agent. We apply the AMAZe to the TPC-H benchmark with the regression indicator removed. The revision agent will then only rely on the provided expert experience to revise the current index recommendation. We can observe that AMAZe's performance only suffers from a small drop. The main reason is that for smaller budgets, the index recommendation with a multi-agent pipeline is straightforward without too many revisions required, as shown in Figure~\ref{fig:pipeline_example}. In addition, the revision agent alone can handle most of the severely regressive indexes by analysing the index utility estimation and learning from the expert experiences.

\begin{table}[t]
\caption{Ablation results conducted on the TPC-H benchmark. For each ablation, the relative workload improvements under 5\%, 10\%, 20\%, and 50\% are averaged.}
\label{tab:ablation}
\resizebox{\linewidth}{!}{
\begin{tabular}
{@{\hskip3pt}l|@{\hskip3pt}c}
\hline
\textbf{Avg. Relative workload improvement}        & (\%)    \\ \hline \hline
\textbf{AMAZe (TPC-H 1x)}  & \textbf{61.33}  \\ 
\hspace{0.2cm} \textbf{- Raw Workload} & 29.52 \\ 
\hspace{0.2cm} \textbf{- One-call Recommendation}  & 11.20 \\ 
\hspace{0.2cm} \textbf{- Raw Workload and One-call Recommendation}  & 10.58 \\ 
\hspace{0.2cm} \textbf{- w/o Regression Indicator}  & 58.21 \\ \hline
\textbf{AMAZe (GPT-3.5-turbo)}  & 60.01  \\
\textbf{AMAZe (QWen3 32b)}  & 60.34  \\ 
\textbf{AMAZe (QWen3 14b)}  & 40.86  \\ 
\textbf{AMAZe (DeepCoder 14b)}  & 40.27  \\ \hline
\textbf{AMAZe (TPC-H 20x)}  &  61.31 \\ \hline
\end{tabular}}
\end{table}

\noindent \textbf{Different LLM Backbones}
\label{sec:84}
To evaluate AMAZe's performance given different LLM backbones, we adopt an API GPT-3.5-turbo~\cite{brown2020language}, and three open-source LLMs, namely Qwen3-32b~\cite{yang2025qwen3technicalreport}, Qwen3-14b and DeepCoder-14b~\cite{deepcoder2025}. DeepCoder 14b is the latest LLM that is specially finetuned on code-related data and achieves SOTA performance in code-related tasks. The results of adopting different LLM backbones are shown in Table~\ref{tab:ablation}. Since AMAZe relies on the reasoning ability of the LLM, we observe that GPT-3.5-turbo and Qwen3-32b have performance comparable to GPT-4o-0806 and outperform Qwen3-14b. Smaller LLMs can also perform index recommendation following AMAZe pipeline, but with a lower performance. It is note-worthy that the DeepCoder, specially fine-tuned on code-related data, underperforms the Qwen3 model with the same parameter size, showing that the coding ability of LLMs is not a must, given the rich information provided by the column candidates. In contrast, the reasoning ability of LLMs is more important.

\noindent \textbf{Different Workload Scales}
In real-world applications, the workloads may always be updating, and such a situation is usually referred to as a dynamic setting. In order to simulate the change of scales, we construct another TPC-H workload of 380 queries named TPC-H 20x, where 20 random queries are generated for each query template. As shown in Table~\ref{tab:ablation}, AMAZe's performance is robust against different workload scales. In addition, since the number of column candidates for the TPC-H 20x is almost the same as TPC-H 1x, the algorithm runtime for TPC-H 20x only increases from an average of 88.80 seconds to 92.91 seconds.

\section{Conclusion and Discussion}
In this work, we present AMAZe, a novel zero-shot index advisor that leverages the reasoning power of large language models through a hierarchical multi-agent pipeline. By representing the workload using column candidates and decomposing the index recommendation task into coordinated sub-steps handled by specialized agents, AMAZe achieves superior performance, strong generalisability, and high efficiency without the need for database-specific training. Extensive experiments across diverse benchmarks demonstrate that AMAZe not only outperforms traditional heuristic and learning-based approaches but also surpasses existing prompt-based LLM methods. 

It is also important to acknowledge the current limitations and outline potential directions for future work. The efficiency of AMAZe may be further improved through the design of more precise and effective instruction sets. Additionally, fine-tuning the agent reasoning pipeline~\cite{ftrag} offers a promising avenue to reduce planning overhead and facilitate deployment on local LLMs. Furthermore, by capitalizing on the advanced reasoning and generalization capabilities of LLMs, the proposed multi-agent framework could be extended to address the index recommendation problem across heterogeneous DBMS environments and diverse SQL dialects. 


\begin{acks}
This work was supported by DAMO Academy through the DAMO Academy Research Intern Program.
\end{acks}


\bibliographystyle{ACM-Reference-Format}
\bibliography{sample}


\begin{thebibliography}{49}


\ifx \showCODEN    \undefined \def \showCODEN     #1{\unskip}     \fi
\ifx \showDOI      \undefined \def \showDOI       #1{#1}\fi
\ifx \showISBNx    \undefined \def \showISBNx     #1{\unskip}     \fi
\ifx \showISBNxiii \undefined \def \showISBNxiii  #1{\unskip}     \fi
\ifx \showISSN     \undefined \def \showISSN      #1{\unskip}     \fi
\ifx \showLCCN     \undefined \def \showLCCN      #1{\unskip}     \fi
\ifx \shownote     \undefined \def \shownote      #1{#1}          \fi
\ifx \showarticletitle \undefined \def \showarticletitle #1{#1}   \fi
\ifx \showURL      \undefined \def \showURL       {\relax}        \fi
\providecommand\bibfield[2]{#2}
\providecommand\bibinfo[2]{#2}
\providecommand\natexlab[1]{#1}
\providecommand\showeprint[2][]{arXiv:#2}

\bibitem[\protect\citeauthoryear{??}{dsb}{[n.d.]}]%
        {dsbgit}
 \bibinfo{year}{[n.d.]}\natexlab{}.
\newblock \bibinfo{title}{DSB Benchmark}.
\newblock \bibinfo{howpublished}{\url{https://github.com/microsoft/dsb/tree/main}}.
\newblock


\bibitem[\protect\citeauthoryear{??}{hyp}{[n.d.]}]%
        {hypopg}
 \bibinfo{year}{[n.d.]}\natexlab{}.
\newblock \bibinfo{title}{HypoPG Extension}.
\newblock \bibinfo{howpublished}{\url{https://github.com/HypoPG/hypopg}}.
\newblock


\bibitem[\protect\citeauthoryear{??}{Pos}{[n.d.]}]%
        {PostgreSQL}
 \bibinfo{year}{[n.d.]}\natexlab{}.
\newblock \bibinfo{title}{PostgreSQL}.
\newblock \bibinfo{howpublished}{\url{https://www.postgresql.org}}.
\newblock


\bibitem[\protect\citeauthoryear{??}{TPC}{[n.d.]}]%
        {TPC-H.tools}
 \bibinfo{year}{[n.d.]}\natexlab{}.
\newblock \bibinfo{title}{TPC-H Toolkit}.
\newblock \bibinfo{howpublished}{\url{https://www.tpc.org/tpc_documents_current_versions/current_specifications5.asp}}.
\newblock


\bibitem[\protect\citeauthoryear{Achiam and et~al.}{Achiam and et~al.}{2023}]%
        {openai2023gpt4}
\bibfield{author}{\bibinfo{person}{Josh Achiam} {and} \bibinfo{person}{Steven~Adler et al.}} \bibinfo{year}{2023}\natexlab{}.
\newblock \showarticletitle{GPT-4 Technical Report}.
\newblock \bibinfo{journal}{\emph{CoRR}}  \bibinfo{volume}{abs/2303.08774} (\bibinfo{date}{March} \bibinfo{year}{2023}).
\newblock
\urldef\tempurl%
\url{https://arxiv.org/abs/2303.08774}
\showURL{%
\tempurl}


\bibitem[\protect\citeauthoryear{Brown, Mann, Ryder, Subbiah, Kaplan, Dhariwal, Neelakantan, Shyam, Sastry, Askell, Agarwal, Herbert-Voss, Krueger, Henighan, Child, Ramesh, Ziegler, Wu, Winter, Hesse, Chen, Sigler, Litwin, Gray, Chess, Clark, Berner, McCandlish, Radford, Sutskever, and Amodei}{Brown et~al\mbox{.}}{2020}]%
        {brown2020language}
\bibfield{author}{\bibinfo{person}{Tom~B. Brown}, \bibinfo{person}{Benjamin Mann}, \bibinfo{person}{Nick Ryder}, \bibinfo{person}{Melanie Subbiah}, \bibinfo{person}{Jared Kaplan}, \bibinfo{person}{Prafulla Dhariwal}, \bibinfo{person}{Arvind Neelakantan}, \bibinfo{person}{Pranav Shyam}, \bibinfo{person}{Girish Sastry}, \bibinfo{person}{Amanda Askell}, \bibinfo{person}{Sandhini Agarwal}, \bibinfo{person}{Ariel Herbert-Voss}, \bibinfo{person}{Gretchen Krueger}, \bibinfo{person}{Tom Henighan}, \bibinfo{person}{Rewon Child}, \bibinfo{person}{Aditya Ramesh}, \bibinfo{person}{Daniel~M. Ziegler}, \bibinfo{person}{Jeffrey Wu}, \bibinfo{person}{Clemens Winter}, \bibinfo{person}{Christopher Hesse}, \bibinfo{person}{Mark Chen}, \bibinfo{person}{Eric Sigler}, \bibinfo{person}{Mateusz Litwin}, \bibinfo{person}{Scott Gray}, \bibinfo{person}{Benjamin Chess}, \bibinfo{person}{Jack Clark}, \bibinfo{person}{Christopher Berner}, \bibinfo{person}{Sam McCandlish}, \bibinfo{person}{Alec Radford}, \bibinfo{person}{Ilya Sutskever},
  {and} \bibinfo{person}{Dario Amodei}.} \bibinfo{year}{2020}\natexlab{}.
\newblock \showarticletitle{Language Models are Few-Shot Learners}.
\newblock \bibinfo{journal}{\emph{CoRR}}  \bibinfo{volume}{abs/2005.14165} (\bibinfo{date}{May} \bibinfo{year}{2020}).
\newblock
\urldef\tempurl%
\url{https://arxiv.org/abs/2005.14165}
\showURL{%
\tempurl}


\bibitem[\protect\citeauthoryear{Brucato, Siddiqui, Wu, Narasayya, and Chaudhuri}{Brucato et~al\mbox{.}}{2024}]%
        {wred}
\bibfield{author}{\bibinfo{person}{Matteo Brucato}, \bibinfo{person}{Tarique Siddiqui}, \bibinfo{person}{Wentao Wu}, \bibinfo{person}{Vivek Narasayya}, {and} \bibinfo{person}{Surajit Chaudhuri}.} \bibinfo{year}{2024}\natexlab{}.
\newblock \showarticletitle{Wred: Workload Reduction for Scalable Index Tuning}.
\newblock \bibinfo{journal}{\emph{Proc. ACM Manag. Data}} \bibinfo{volume}{2}, \bibinfo{number}{1}, Article \bibinfo{articleno}{50} (\bibinfo{date}{March} \bibinfo{year}{2024}), \bibinfo{numpages}{26}~pages.
\newblock
\urldef\tempurl%
\url{https://doi.org/10.1145/3639305}
\showDOI{\tempurl}


\bibitem[\protect\citeauthoryear{Bruno and Chaudhuri}{Bruno and Chaudhuri}{2005}]%
        {10.1145/1066157.1066184}
\bibfield{author}{\bibinfo{person}{Nicolas Bruno} {and} \bibinfo{person}{Surajit Chaudhuri}.} \bibinfo{year}{2005}\natexlab{}.
\newblock \showarticletitle{Automatic physical database tuning: a relaxation-based approach}. In \bibinfo{booktitle}{\emph{Proceedings of the 2005 ACM SIGMOD International Conference on Management of Data}} (Baltimore, Maryland) \emph{(\bibinfo{series}{SIGMOD '05})}. \bibinfo{publisher}{Association for Computing Machinery}, \bibinfo{address}{New York, NY, USA}, \bibinfo{pages}{227–238}.
\newblock
\showISBNx{1595930604}
\urldef\tempurl%
\url{https://doi.org/10.1145/1066157.1066184}
\showDOI{\tempurl}


\bibitem[\protect\citeauthoryear{Chang, Zhang, Li, Miao, Qin, and Cui}{Chang et~al\mbox{.}}{2024}]%
        {mfix}
\bibfield{author}{\bibinfo{person}{Zhuo Chang}, \bibinfo{person}{Xinyi Zhang}, \bibinfo{person}{Yang Li}, \bibinfo{person}{Xupeng Miao}, \bibinfo{person}{Yanzhao Qin}, {and} \bibinfo{person}{Bin Cui}.} \bibinfo{year}{2024}\natexlab{}.
\newblock \showarticletitle{MFIX: An Efficient and Reliable Index Advisor via Multi-Fidelity Bayesian Optimization}. In \bibinfo{booktitle}{\emph{2024 IEEE 40th International Conference on Data Engineering (ICDE)}}. \bibinfo{pages}{4343--4356}.
\newblock
\urldef\tempurl%
\url{https://doi.org/10.1109/ICDE60146.2024.00331}
\showDOI{\tempurl}


\bibitem[\protect\citeauthoryear{Chaudhuri and Narasayya}{Chaudhuri and Narasayya}{2020}]%
        {chaudhuri2020anytime}
\bibfield{author}{\bibinfo{person}{Surajit Chaudhuri} {and} \bibinfo{person}{Vivek Narasayya}.} \bibinfo{year}{2020}\natexlab{}.
\newblock \bibinfo{title}{Anytime Algorithm of Database Tuning Advisor for Microsoft SQL Server}.  (\bibinfo{date}{June} \bibinfo{year}{2020}).
\newblock
\urldef\tempurl%
\url{https://www.microsoft.com/en-us/research/publication/anytime-algorithm-of-database-tuning-advisor-for-microsoft-sql-server/}
\showURL{%
\tempurl}


\bibitem[\protect\citeauthoryear{Chaudhuri and Narasayya}{Chaudhuri and Narasayya}{1997}]%
        {10.5555/645923.673646}
\bibfield{author}{\bibinfo{person}{Surajit Chaudhuri} {and} \bibinfo{person}{Vivek~R. Narasayya}.} \bibinfo{year}{1997}\natexlab{}.
\newblock \showarticletitle{An Efficient Cost-Driven Index Selection Tool for Microsoft SQL Server}. In \bibinfo{booktitle}{\emph{Proceedings of the 23rd International Conference on Very Large Data Bases}} \emph{(\bibinfo{series}{VLDB '97})}. \bibinfo{publisher}{Morgan Kaufmann Publishers Inc.}, \bibinfo{address}{San Francisco, CA, USA}, \bibinfo{pages}{146–155}.
\newblock
\showISBNx{1558604707}


\bibitem[\protect\citeauthoryear{Choenni, Blanken, and Chang}{Choenni et~al\mbox{.}}{1993}]%
        {315323}
\bibfield{author}{\bibinfo{person}{S. Choenni}, \bibinfo{person}{H. Blanken}, {and} \bibinfo{person}{T. Chang}.} \bibinfo{year}{1993}\natexlab{}.
\newblock \showarticletitle{Index selection in relational databases}. In \bibinfo{booktitle}{\emph{Proceedings of ICCI'93: 5th International Conference on Computing and Information}}. \bibinfo{pages}{491--496}.
\newblock
\urldef\tempurl%
\url{https://doi.org/10.1109/ICCI.1993.315323}
\showDOI{\tempurl}


\bibitem[\protect\citeauthoryear{Ding, Chaudhuri, Gehrke, and Narasayya}{Ding et~al\mbox{.}}{2021}]%
        {ding2021dsb}
\bibfield{author}{\bibinfo{person}{Bailu Ding}, \bibinfo{person}{Surajit Chaudhuri}, \bibinfo{person}{Johannes Gehrke}, {and} \bibinfo{person}{Vivek Narasayya}.} \bibinfo{year}{2021}\natexlab{}.
\newblock \showarticletitle{DSB: A Decision Support Benchmark for Workload-Driven and Traditional Database Systems}. In \bibinfo{booktitle}{\emph{VLDB 2022}}.
\newblock
\urldef\tempurl%
\url{https://www.microsoft.com/en-us/research/publication/dsb-a-decision-support-benchmark-for-workload-driven-and-traditional-database-systems/}
\showURL{%
\tempurl}


\bibitem[\protect\citeauthoryear{Ding, Das, Marcus, Wu, Chaudhuri, and Narasayya}{Ding et~al\mbox{.}}{2019}]%
        {aimeetsai}
\bibfield{author}{\bibinfo{person}{Bailu Ding}, \bibinfo{person}{Sudipto Das}, \bibinfo{person}{Ryan Marcus}, \bibinfo{person}{Wentao Wu}, \bibinfo{person}{Surajit Chaudhuri}, {and} \bibinfo{person}{Vivek~R. Narasayya}.} \bibinfo{year}{2019}\natexlab{}.
\newblock \showarticletitle{AI Meets AI: Leveraging Query Executions to Improve Index Recommendations}. In \bibinfo{booktitle}{\emph{Proceedings of the 2019 International Conference on Management of Data}} (Amsterdam, Netherlands) \emph{(\bibinfo{series}{SIGMOD '19})}. \bibinfo{publisher}{Association for Computing Machinery}, \bibinfo{address}{New York, NY, USA}, \bibinfo{pages}{1241–1258}.
\newblock
\showISBNx{9781450356435}
\urldef\tempurl%
\url{https://doi.org/10.1145/3299869.3324957}
\showDOI{\tempurl}


\bibitem[\protect\citeauthoryear{Jha, Carvalho, Liang, Du, Kleiman-Weiner, and Jaques}{Jha et~al\mbox{.}}{2025}]%
        {jha2025crossenvironment}
\bibfield{author}{\bibinfo{person}{Kunal Jha}, \bibinfo{person}{Wilka Carvalho}, \bibinfo{person}{Yancheng Liang}, \bibinfo{person}{Simon~Shaolei Du}, \bibinfo{person}{Max Kleiman-Weiner}, {and} \bibinfo{person}{Natasha Jaques}.} \bibinfo{year}{2025}\natexlab{}.
\newblock \showarticletitle{Cross-environment Cooperation Enables Zero-shot Multi-agent Coordination}. In \bibinfo{booktitle}{\emph{Forty-second International Conference on Machine Learning}}.
\newblock
\urldef\tempurl%
\url{https://openreview.net/forum?id=zBBYsVGKuB}
\showURL{%
\tempurl}


\bibitem[\protect\citeauthoryear{Ji, Lee, Frieske, Yu, Su, Xu, Ishii, Bang, Madotto, and Fung}{Ji et~al\mbox{.}}{2023}]%
        {Ji_2023}
\bibfield{author}{\bibinfo{person}{Ziwei Ji}, \bibinfo{person}{Nayeon Lee}, \bibinfo{person}{Rita Frieske}, \bibinfo{person}{Tiezheng Yu}, \bibinfo{person}{Dan Su}, \bibinfo{person}{Yan Xu}, \bibinfo{person}{Etsuko Ishii}, \bibinfo{person}{Ye~Jin Bang}, \bibinfo{person}{Andrea Madotto}, {and} \bibinfo{person}{Pascale Fung}.} \bibinfo{year}{2023}\natexlab{}.
\newblock \showarticletitle{Survey of Hallucination in Natural Language Generation}.
\newblock \bibinfo{journal}{\emph{Comput. Surveys}} \bibinfo{volume}{55}, \bibinfo{number}{12} (\bibinfo{date}{March} \bibinfo{year}{2023}), \bibinfo{pages}{1–38}.
\newblock
\showISSN{1557-7341}
\urldef\tempurl%
\url{https://doi.org/10.1145/3571730}
\showDOI{\tempurl}


\bibitem[\protect\citeauthoryear{Kossmann, Halfpap, Jankrift, and Schlosser}{Kossmann et~al\mbox{.}}{2020}]%
        {magicmirror}
\bibfield{author}{\bibinfo{person}{Jan Kossmann}, \bibinfo{person}{Stefan Halfpap}, \bibinfo{person}{Marcel Jankrift}, {and} \bibinfo{person}{Rainer Schlosser}.} \bibinfo{year}{2020}\natexlab{}.
\newblock \showarticletitle{Magic mirror in my hand, which is the best in the land? an experimental evaluation of index selection algorithms}.
\newblock \bibinfo{journal}{\emph{Proc. VLDB Endow.}} \bibinfo{volume}{13}, \bibinfo{number}{12} (\bibinfo{date}{July} \bibinfo{year}{2020}), \bibinfo{pages}{2382–2395}.
\newblock
\showISSN{2150-8097}
\urldef\tempurl%
\url{https://doi.org/10.14778/3407790.3407832}
\showDOI{\tempurl}


\bibitem[\protect\citeauthoryear{Kossmann, Kastius, and Schlosser}{Kossmann et~al\mbox{.}}{2022}]%
        {swirl}
\bibfield{author}{\bibinfo{person}{Jan Kossmann}, \bibinfo{person}{Alexander Kastius}, {and} \bibinfo{person}{Rainer Schlosser}.} \bibinfo{year}{2022}\natexlab{}.
\newblock \showarticletitle{SWIRL: Selection of Workload-aware Indexes using Reinforcement Learning}. In \bibinfo{booktitle}{\emph{Proceedings of the 25th International Conference on Extending Database Technology (EDBT)}}. \bibinfo{pages}{2:155– 2:168}.
\newblock


\bibitem[\protect\citeauthoryear{Lan, Bao, and Peng}{Lan et~al\mbox{.}}{2020}]%
        {10.1145/3340531.3412106}
\bibfield{author}{\bibinfo{person}{Hai Lan}, \bibinfo{person}{Zhifeng Bao}, {and} \bibinfo{person}{Yuwei Peng}.} \bibinfo{year}{2020}\natexlab{}.
\newblock \showarticletitle{An Index Advisor Using Deep Reinforcement Learning}. In \bibinfo{booktitle}{\emph{Proceedings of the 29th ACM International Conference on Information \& Knowledge Management}} (Virtual Event, Ireland) \emph{(\bibinfo{series}{CIKM '20})}. \bibinfo{publisher}{Association for Computing Machinery}, \bibinfo{address}{New York, NY, USA}, \bibinfo{pages}{2105–2108}.
\newblock
\showISBNx{9781450368599}
\urldef\tempurl%
\url{https://doi.org/10.1145/3340531.3412106}
\showDOI{\tempurl}


\bibitem[\protect\citeauthoryear{Leis, Gubichev, Mirchev, Boncz, Kemper, and Neumann}{Leis et~al\mbox{.}}{2015}]%
        {Leis2015HowGA}
\bibfield{author}{\bibinfo{person}{Viktor Leis}, \bibinfo{person}{Andrey Gubichev}, \bibinfo{person}{Atanas Mirchev}, \bibinfo{person}{Peter~A. Boncz}, \bibinfo{person}{Alfons Kemper}, {and} \bibinfo{person}{Thomas Neumann}.} \bibinfo{year}{2015}\natexlab{}.
\newblock \showarticletitle{How Good Are Query Optimizers, Really?}
\newblock \bibinfo{journal}{\emph{Proc. VLDB Endow.}}  \bibinfo{volume}{9} (\bibinfo{year}{2015}), \bibinfo{pages}{204--215}.
\newblock
\urldef\tempurl%
\url{https://api.semanticscholar.org/CorpusID:7953847}
\showURL{%
\tempurl}


\bibitem[\protect\citeauthoryear{Li, Xie, Li, Tsung, Ding, and Li}{Li et~al\mbox{.}}{2025}]%
        {li2025agentoriented}
\bibfield{author}{\bibinfo{person}{Ao Li}, \bibinfo{person}{Yuexiang Xie}, \bibinfo{person}{Songze Li}, \bibinfo{person}{Fugee Tsung}, \bibinfo{person}{Bolin Ding}, {and} \bibinfo{person}{Yaliang Li}.} \bibinfo{year}{2025}\natexlab{}.
\newblock \showarticletitle{Agent-Oriented Planning in Multi-Agent Systems}. In \bibinfo{booktitle}{\emph{The Thirteenth International Conference on Learning Representations}}.
\newblock
\urldef\tempurl%
\url{https://openreview.net/forum?id=EqcLAU6gyU}
\showURL{%
\tempurl}


\bibitem[\protect\citeauthoryear{Liu, Singh, Liu, Payani, and Zheng}{Liu et~al\mbox{.}}{2025}]%
        {liu2025towards}
\bibfield{author}{\bibinfo{person}{Yuchi Liu}, \bibinfo{person}{Jaskirat Singh}, \bibinfo{person}{Gaowen Liu}, \bibinfo{person}{Ali Payani}, {and} \bibinfo{person}{Liang Zheng}.} \bibinfo{year}{2025}\natexlab{}.
\newblock \showarticletitle{Towards Hierarchical Multi-Agent Workflows for Zero-Shot Prompt Optimization}. In \bibinfo{booktitle}{\emph{Workshop on Reasoning and Planning for Large Language Models}}.
\newblock
\urldef\tempurl%
\url{https://openreview.net/forum?id=RVvXOrP2qm}
\showURL{%
\tempurl}


\bibitem[\protect\citeauthoryear{Luo, Tan, Huang, Patel, Ariyak, Wu, Shi, Xin, Cai, Weber, Zhang, Li, Popa, and Stoica}{Luo et~al\mbox{.}}{2025}]%
        {deepcoder2025}
\bibfield{author}{\bibinfo{person}{Michael Luo}, \bibinfo{person}{Sijun Tan}, \bibinfo{person}{Roy Huang}, \bibinfo{person}{Ameen Patel}, \bibinfo{person}{Alpay Ariyak}, \bibinfo{person}{Qingyang Wu}, \bibinfo{person}{Xiaoxiang Shi}, \bibinfo{person}{Rachel Xin}, \bibinfo{person}{Colin Cai}, \bibinfo{person}{Maurice Weber}, \bibinfo{person}{Ce Zhang}, \bibinfo{person}{Li~Erran Li}, \bibinfo{person}{Raluca~Ada Popa}, {and} \bibinfo{person}{Ion Stoica}.} \bibinfo{year}{2025}\natexlab{}.
\newblock \bibinfo{title}{DeepCoder: A Fully Open-Source 14B Coder at O3-mini Level}.
\newblock \bibinfo{howpublished}{\url{https://pretty-radio-b75.notion.site/DeepCoder-A-Fully-Open-Source-14B-Coder-at-O3-mini-Level-1cf81902c14680b3bee5eb349a512a51}}.
\newblock
\newblock
\shownote{Notion Blog.}


\bibitem[\protect\citeauthoryear{Maas, Daly, Pham, Huang, Ng, and Potts}{Maas et~al\mbox{.}}{2011}]%
        {maas-EtAl:2011:ACL-HLT2011}
\bibfield{author}{\bibinfo{person}{Andrew~L. Maas}, \bibinfo{person}{Raymond~E. Daly}, \bibinfo{person}{Peter~T. Pham}, \bibinfo{person}{Dan Huang}, \bibinfo{person}{Andrew~Y. Ng}, {and} \bibinfo{person}{Christopher Potts}.} \bibinfo{year}{2011}\natexlab{}.
\newblock \showarticletitle{Learning Word Vectors for Sentiment Analysis}. In \bibinfo{booktitle}{\emph{Proceedings of the 49th Annual Meeting of the Association for Computational Linguistics: Human Language Technologies}}. \bibinfo{publisher}{Association for Computational Linguistics}, \bibinfo{address}{Portland, Oregon, USA}, \bibinfo{pages}{142--150}.
\newblock
\urldef\tempurl%
\url{http://www.aclweb.org/anthology/P11-1015}
\showURL{%
\tempurl}


\bibitem[\protect\citeauthoryear{Paludo~Licks, Colleoni~Couto, de~F\'{a}tima~Miehe, de~Paris, Dubugras~Ruiz, and Meneguzzi}{Paludo~Licks et~al\mbox{.}}{2020}]%
        {10.1007/s10489-020-01674-8}
\bibfield{author}{\bibinfo{person}{Gabriel Paludo~Licks}, \bibinfo{person}{Julia Colleoni~Couto}, \bibinfo{person}{Priscilla de F\'{a}tima~Miehe}, \bibinfo{person}{Renata de Paris}, \bibinfo{person}{Duncan Dubugras~Ruiz}, {and} \bibinfo{person}{Felipe Meneguzzi}.} \bibinfo{year}{2020}\natexlab{}.
\newblock \showarticletitle{SmartIX: A database indexing agent based on reinforcement learning}.
\newblock \bibinfo{journal}{\emph{Applied Intelligence}} \bibinfo{volume}{50}, \bibinfo{number}{8} (\bibinfo{date}{Aug.} \bibinfo{year}{2020}), \bibinfo{pages}{2575–2588}.
\newblock
\showISSN{0924-669X}
\urldef\tempurl%
\url{https://doi.org/10.1007/s10489-020-01674-8}
\showDOI{\tempurl}


\bibitem[\protect\citeauthoryear{Perera, Oetomo, Rubinstein, and Borovica{-}Gajic}{Perera et~al\mbox{.}}{2021a}]%
        {DBLP:conf/icde/PereraORB21}
\bibfield{author}{\bibinfo{person}{R.~Malinga Perera}, \bibinfo{person}{Bastian Oetomo}, \bibinfo{person}{Benjamin I.~P. Rubinstein}, {and} \bibinfo{person}{Renata Borovica{-}Gajic}.} \bibinfo{year}{2021}\natexlab{a}.
\newblock \showarticletitle{{DBA} bandits: Self-driving index tuning under ad-hoc, analytical workloads with safety guarantees}. In \bibinfo{booktitle}{\emph{37th {IEEE} International Conference on Data Engineering, {ICDE} 2021, Chania, Greece, April 19-22, 2021}}. \bibinfo{publisher}{{IEEE}}, \bibinfo{pages}{600--611}.
\newblock
\urldef\tempurl%
\url{https://doi.org/10.1109/ICDE51399.2021.00058}
\showDOI{\tempurl}


\bibitem[\protect\citeauthoryear{Perera, Oetomo, Rubinstein, and Borovica-Gajic}{Perera et~al\mbox{.}}{2021b}]%
        {DBAbandits}
\bibfield{author}{\bibinfo{person}{R.~Malinga Perera}, \bibinfo{person}{Bastian Oetomo}, \bibinfo{person}{Benjamin I.~P. Rubinstein}, {and} \bibinfo{person}{Renata Borovica-Gajic}.} \bibinfo{year}{2021}\natexlab{b}.
\newblock \showarticletitle{DBA bandits: Self-driving index tuning under ad-hoc, analytical workloads with safety guarantees}. In \bibinfo{booktitle}{\emph{2021 IEEE 37th International Conference on Data Engineering (ICDE)}}. \bibinfo{pages}{600--611}.
\newblock
\urldef\tempurl%
\url{https://doi.org/10.1109/ICDE51399.2021.00058}
\showDOI{\tempurl}


\bibitem[\protect\citeauthoryear{Reimers and Gurevych}{Reimers and Gurevych}{2019}]%
        {reimers-2019-sentence-bert}
\bibfield{author}{\bibinfo{person}{Nils Reimers} {and} \bibinfo{person}{Iryna Gurevych}.} \bibinfo{year}{2019}\natexlab{}.
\newblock \showarticletitle{Sentence-BERT: Sentence Embeddings using Siamese BERT-Networks}. In \bibinfo{booktitle}{\emph{Proceedings of the 2019 Conference on Empirical Methods in Natural Language Processing}}. \bibinfo{publisher}{Association for Computational Linguistics}.
\newblock
\urldef\tempurl%
\url{https://arxiv.org/abs/1908.10084}
\showURL{%
\tempurl}


\bibitem[\protect\citeauthoryear{Sadri, Gruenwald, and Lead}{Sadri et~al\mbox{.}}{2020a}]%
        {DRLindex}
\bibfield{author}{\bibinfo{person}{Zahra Sadri}, \bibinfo{person}{Le Gruenwald}, {and} \bibinfo{person}{Eleazar Lead}.} \bibinfo{year}{2020}\natexlab{a}.
\newblock \showarticletitle{DRLindex: deep reinforcement learning index advisor for a cluster database}. In \bibinfo{booktitle}{\emph{Proceedings of the 24th Symposium on International Database Engineering \& Applications}} (Seoul, Republic of Korea) \emph{(\bibinfo{series}{IDEAS '20})}. \bibinfo{publisher}{Association for Computing Machinery}, \bibinfo{address}{New York, NY, USA}, Article \bibinfo{articleno}{11}, \bibinfo{numpages}{8}~pages.
\newblock
\showISBNx{9781450375030}
\urldef\tempurl%
\url{https://doi.org/10.1145/3410566.3410603}
\showDOI{\tempurl}


\bibitem[\protect\citeauthoryear{Sadri, Gruenwald, and Leal}{Sadri et~al\mbox{.}}{2020b}]%
        {9094124}
\bibfield{author}{\bibinfo{person}{Zahra Sadri}, \bibinfo{person}{Le Gruenwald}, {and} \bibinfo{person}{Eleazar Leal}.} \bibinfo{year}{2020}\natexlab{b}.
\newblock \showarticletitle{Online Index Selection Using Deep Reinforcement Learning for a Cluster Database}. In \bibinfo{booktitle}{\emph{2020 IEEE 36th International Conference on Data Engineering Workshops (ICDEW)}}. \bibinfo{pages}{158--161}.
\newblock
\urldef\tempurl%
\url{https://doi.org/10.1109/ICDEW49219.2020.00035}
\showDOI{\tempurl}


\bibitem[\protect\citeauthoryear{Schlosser, Kossmann, and Boissier}{Schlosser et~al\mbox{.}}{2019}]%
        {extend}
\bibfield{author}{\bibinfo{person}{Rainer Schlosser}, \bibinfo{person}{Jan Kossmann}, {and} \bibinfo{person}{Martin Boissier}.} \bibinfo{year}{2019}\natexlab{}.
\newblock \showarticletitle{Efficient Scalable Multi-attribute Index Selection Using Recursive Strategies}. In \bibinfo{booktitle}{\emph{2019 IEEE 35th International Conference on Data Engineering (ICDE)}}. \bibinfo{pages}{1238--1249}.
\newblock
\urldef\tempurl%
\url{https://doi.org/10.1109/ICDE.2019.00113}
\showDOI{\tempurl}


\bibitem[\protect\citeauthoryear{Schulman, Wolski, Dhariwal, Radford, and Klimov}{Schulman et~al\mbox{.}}{2017}]%
        {ppo}
\bibfield{author}{\bibinfo{person}{John Schulman}, \bibinfo{person}{Filip Wolski}, \bibinfo{person}{Prafulla Dhariwal}, \bibinfo{person}{Alec Radford}, {and} \bibinfo{person}{Oleg Klimov}.} \bibinfo{year}{2017}\natexlab{}.
\newblock \showarticletitle{Proximal Policy Optimization Algorithms}.
\newblock \bibinfo{journal}{\emph{CoRR}}  \bibinfo{volume}{abs/1707.06347} (\bibinfo{date}{July} \bibinfo{year}{2017}).
\newblock
\urldef\tempurl%
\url{https://arxiv.org/abs/1707.06347}
\showURL{%
\tempurl}


\bibitem[\protect\citeauthoryear{Shi, Cong, and Li}{Shi et~al\mbox{.}}{2022}]%
        {jia}
\bibfield{author}{\bibinfo{person}{Jiachen Shi}, \bibinfo{person}{Gao Cong}, {and} \bibinfo{person}{Xiao-Li Li}.} \bibinfo{year}{2022}\natexlab{}.
\newblock \showarticletitle{Learned Index Benefits: Machine Learning Based Index Performance Estimation}.
\newblock \bibinfo{journal}{\emph{Proc. VLDB Endow.}} \bibinfo{volume}{15}, \bibinfo{number}{13} (\bibinfo{date}{Sept.} \bibinfo{year}{2022}), \bibinfo{pages}{3950–3962}.
\newblock
\showISSN{2150-8097}
\urldef\tempurl%
\url{https://doi.org/10.14778/3565838.3565848}
\showDOI{\tempurl}


\bibitem[\protect\citeauthoryear{Siddiqui, Jo, Wu, Wang, Narasayya, and Chaudhuri}{Siddiqui et~al\mbox{.}}{2022a}]%
        {isum}
\bibfield{author}{\bibinfo{person}{Tarique Siddiqui}, \bibinfo{person}{Saehan Jo}, \bibinfo{person}{Wentao Wu}, \bibinfo{person}{Chi Wang}, \bibinfo{person}{Vivek Narasayya}, {and} \bibinfo{person}{Surajit Chaudhuri}.} \bibinfo{year}{2022}\natexlab{a}.
\newblock \showarticletitle{ISUM: Efficiently Compressing Large and Complex Workloads for Scalable Index Tuning}. In \bibinfo{booktitle}{\emph{Proceedings of the 2022 International Conference on Management of Data}} (Philadelphia, PA, USA) \emph{(\bibinfo{series}{SIGMOD '22})}. \bibinfo{publisher}{Association for Computing Machinery}, \bibinfo{address}{New York, NY, USA}, \bibinfo{pages}{660–673}.
\newblock
\showISBNx{9781450392495}
\urldef\tempurl%
\url{https://doi.org/10.1145/3514221.3526152}
\showDOI{\tempurl}


\bibitem[\protect\citeauthoryear{Siddiqui, Wu, Narasayya, and Chaudhuri}{Siddiqui et~al\mbox{.}}{2022b}]%
        {distill}
\bibfield{author}{\bibinfo{person}{Tarique Siddiqui}, \bibinfo{person}{Wentao Wu}, \bibinfo{person}{Vivek Narasayya}, {and} \bibinfo{person}{Surajit Chaudhuri}.} \bibinfo{year}{2022}\natexlab{b}.
\newblock \showarticletitle{DISTILL: low-overhead data-driven techniques for filtering and costing indexes for scalable index tuning}.
\newblock \bibinfo{journal}{\emph{Proc. VLDB Endow.}} \bibinfo{volume}{15}, \bibinfo{number}{10} (\bibinfo{date}{June} \bibinfo{year}{2022}), \bibinfo{pages}{2019–2031}.
\newblock
\showISSN{2150-8097}
\urldef\tempurl%
\url{https://doi.org/10.14778/3547305.3547309}
\showDOI{\tempurl}


\bibitem[\protect\citeauthoryear{Skelley}{Skelley}{2000}]%
        {10.5555/846219.847390}
\bibfield{author}{\bibinfo{person}{Alan Skelley}.} \bibinfo{year}{2000}\natexlab{}.
\newblock \showarticletitle{DB2 Advisor: An Optimizer Smart Enough to Recommend its own Indexes}. In \bibinfo{booktitle}{\emph{Proceedings of the 16th International Conference on Data Engineering}} \emph{(\bibinfo{series}{ICDE '00})}. \bibinfo{publisher}{IEEE Computer Society}, \bibinfo{address}{USA}, \bibinfo{pages}{101}.
\newblock
\showISBNx{0769505066}


\bibitem[\protect\citeauthoryear{Wang, Chen, Hu, Yang, Liu, Shen, Wei, Zhang, Gu, Zhou, Pan, Zhang, and Chen}{Wang et~al\mbox{.}}{2024a}]%
        {ftrag}
\bibfield{author}{\bibinfo{person}{Junjie Wang}, \bibinfo{person}{Mingyang Chen}, \bibinfo{person}{Binbin Hu}, \bibinfo{person}{Dan Yang}, \bibinfo{person}{Ziqi Liu}, \bibinfo{person}{Yue Shen}, \bibinfo{person}{Peng Wei}, \bibinfo{person}{Zhiqiang Zhang}, \bibinfo{person}{Jinjie Gu}, \bibinfo{person}{Jun Zhou}, \bibinfo{person}{Jeff~Z. Pan}, \bibinfo{person}{Wen Zhang}, {and} \bibinfo{person}{Huajun Chen}.} \bibinfo{year}{2024}\natexlab{a}.
\newblock \showarticletitle{Learning to Plan for Retrieval-Augmented Large Language Models from Knowledge Graphs}. In \bibinfo{booktitle}{\emph{EMNLP (Findings)}}. \bibinfo{pages}{7813--7835}.
\newblock
\urldef\tempurl%
\url{https://aclanthology.org/2024.findings-emnlp.459}
\showURL{%
\tempurl}


\bibitem[\protect\citeauthoryear{Wang and Yoneki}{Wang and Yoneki}{2024}]%
        {ia2}
\bibfield{author}{\bibinfo{person}{Taiyi Wang} {and} \bibinfo{person}{Eiko Yoneki}.} \bibinfo{year}{2024}\natexlab{}.
\newblock \showarticletitle{IA2: Leveraging Instance-Aware Index Advisor with Reinforcement Learning for Diverse Workloads}. In \bibinfo{booktitle}{\emph{Proceedings of the 4th Workshop on Machine Learning and Systems}} (Athens, Greece) \emph{(\bibinfo{series}{EuroMLSys '24})}. \bibinfo{publisher}{Association for Computing Machinery}, \bibinfo{address}{New York, NY, USA}, \bibinfo{pages}{10–17}.
\newblock
\showISBNx{9798400705410}
\urldef\tempurl%
\url{https://doi.org/10.1145/3642970.3655839}
\showDOI{\tempurl}


\bibitem[\protect\citeauthoryear{Wang, Liu, Lin, Bao, Li, and Wang}{Wang et~al\mbox{.}}{2024b}]%
        {balance}
\bibfield{author}{\bibinfo{person}{Zijia Wang}, \bibinfo{person}{Haoran Liu}, \bibinfo{person}{Chen Lin}, \bibinfo{person}{Zhifeng Bao}, \bibinfo{person}{Guoliang Li}, {and} \bibinfo{person}{Tianqing Wang}.} \bibinfo{year}{2024}\natexlab{b}.
\newblock \showarticletitle{Leveraging Dynamic and Heterogeneous Workload Knowledge to Boost the Performance of Index Advisors}.
\newblock \bibinfo{journal}{\emph{Proc. VLDB Endow.}} \bibinfo{volume}{17}, \bibinfo{number}{7} (\bibinfo{date}{March} \bibinfo{year}{2024}), \bibinfo{pages}{1642–1654}.
\newblock
\showISSN{2150-8097}
\urldef\tempurl%
\url{https://doi.org/10.14778/3654621.3654631}
\showDOI{\tempurl}


\bibitem[\protect\citeauthoryear{Wei, Wei, Tay, Tran, Webson, Lu, Chen, Liu, Huang, Zhou, and Ma}{Wei et~al\mbox{.}}{2023}]%
        {wei2023larger}
\bibfield{author}{\bibinfo{person}{Jerry Wei}, \bibinfo{person}{Jason Wei}, \bibinfo{person}{Yi Tay}, \bibinfo{person}{Dustin Tran}, \bibinfo{person}{Albert Webson}, \bibinfo{person}{Yifeng Lu}, \bibinfo{person}{Xinyun Chen}, \bibinfo{person}{Hanxiao Liu}, \bibinfo{person}{Da Huang}, \bibinfo{person}{Denny Zhou}, {and} \bibinfo{person}{Tengyu Ma}.} \bibinfo{year}{2023}\natexlab{}.
\newblock \showarticletitle{Larger language models do in-context learning differently}.
\newblock \bibinfo{journal}{\emph{CoRR}}  \bibinfo{volume}{abs/2303.03846} (\bibinfo{date}{March} \bibinfo{year}{2023}).
\newblock
\urldef\tempurl%
\url{https://arxiv.org/abs/2303.03846}
\showURL{%
\tempurl}


\bibitem[\protect\citeauthoryear{Wu, Wang, Siddiqui, Wang, Narasayya, Chaudhuri, and Bernstein}{Wu et~al\mbox{.}}{2022}]%
        {budgetaware}
\bibfield{author}{\bibinfo{person}{Wentao Wu}, \bibinfo{person}{Chi Wang}, \bibinfo{person}{Tarique Siddiqui}, \bibinfo{person}{Junxiong Wang}, \bibinfo{person}{Vivek Narasayya}, \bibinfo{person}{Surajit Chaudhuri}, {and} \bibinfo{person}{Philip~A. Bernstein}.} \bibinfo{year}{2022}\natexlab{}.
\newblock \showarticletitle{Budget-aware Index Tuning with Reinforcement Learning}. In \bibinfo{booktitle}{\emph{Proceedings of the 2022 International Conference on Management of Data}} (Philadelphia, PA, USA) \emph{(\bibinfo{series}{SIGMOD '22})}. \bibinfo{publisher}{Association for Computing Machinery}, \bibinfo{address}{New York, NY, USA}, \bibinfo{pages}{1528–1541}.
\newblock
\showISBNx{9781450392495}
\urldef\tempurl%
\url{https://doi.org/10.1145/3514221.3526128}
\showDOI{\tempurl}


\bibitem[\protect\citeauthoryear{Yang, Li, Yang, Zhang, Hui, Zheng, Yu, Gao, Huang, Lv, Zheng, Liu, Zhou, Huang, Hu, Ge, Wei, Lin, Tang, Yang, Tu, Zhang, Yang, Yang, Zhou, Zhou, Lin, Dang, Bao, Yang, Yu, Deng, Li, Xue, Li, Zhang, Wang, Zhu, Men, Gao, Liu, Luo, Li, Tang, Yin, Ren, Wang, Zhang, Ren, Fan, Su, Zhang, Zhang, Wan, Liu, Wang, Cui, Zhang, Zhou, and Qiu}{Yang et~al\mbox{.}}{2025}]%
        {yang2025qwen3technicalreport}
\bibfield{author}{\bibinfo{person}{An Yang}, \bibinfo{person}{Anfeng Li}, \bibinfo{person}{Baosong Yang}, \bibinfo{person}{Beichen Zhang}, \bibinfo{person}{Binyuan Hui}, \bibinfo{person}{Bo Zheng}, \bibinfo{person}{Bowen Yu}, \bibinfo{person}{Chang Gao}, \bibinfo{person}{Chengen Huang}, \bibinfo{person}{Chenxu Lv}, \bibinfo{person}{Chujie Zheng}, \bibinfo{person}{Dayiheng Liu}, \bibinfo{person}{Fan Zhou}, \bibinfo{person}{Fei Huang}, \bibinfo{person}{Feng Hu}, \bibinfo{person}{Hao Ge}, \bibinfo{person}{Haoran Wei}, \bibinfo{person}{Huan Lin}, \bibinfo{person}{Jialong Tang}, \bibinfo{person}{Jian Yang}, \bibinfo{person}{Jianhong Tu}, \bibinfo{person}{Jianwei Zhang}, \bibinfo{person}{Jianxin Yang}, \bibinfo{person}{Jiaxi Yang}, \bibinfo{person}{Jing Zhou}, \bibinfo{person}{Jingren Zhou}, \bibinfo{person}{Junyang Lin}, \bibinfo{person}{Kai Dang}, \bibinfo{person}{Keqin Bao}, \bibinfo{person}{Kexin Yang}, \bibinfo{person}{Le Yu}, \bibinfo{person}{Lianghao Deng}, \bibinfo{person}{Mei Li}, \bibinfo{person}{Mingfeng
  Xue}, \bibinfo{person}{Mingze Li}, \bibinfo{person}{Pei Zhang}, \bibinfo{person}{Peng Wang}, \bibinfo{person}{Qin Zhu}, \bibinfo{person}{Rui Men}, \bibinfo{person}{Ruize Gao}, \bibinfo{person}{Shixuan Liu}, \bibinfo{person}{Shuang Luo}, \bibinfo{person}{Tianhao Li}, \bibinfo{person}{Tianyi Tang}, \bibinfo{person}{Wenbiao Yin}, \bibinfo{person}{Xingzhang Ren}, \bibinfo{person}{Xinyu Wang}, \bibinfo{person}{Xinyu Zhang}, \bibinfo{person}{Xuancheng Ren}, \bibinfo{person}{Yang Fan}, \bibinfo{person}{Yang Su}, \bibinfo{person}{Yichang Zhang}, \bibinfo{person}{Yinger Zhang}, \bibinfo{person}{Yu Wan}, \bibinfo{person}{Yuqiong Liu}, \bibinfo{person}{Zekun Wang}, \bibinfo{person}{Zeyu Cui}, \bibinfo{person}{Zhenru Zhang}, \bibinfo{person}{Zhipeng Zhou}, {and} \bibinfo{person}{Zihan Qiu}.} \bibinfo{year}{2025}\natexlab{}.
\newblock \showarticletitle{Qwen3 Technical Report}.
\newblock \bibinfo{journal}{\emph{CoRR}}  \bibinfo{volume}{abs/2505.09388} (\bibinfo{date}{May} \bibinfo{year}{2025}).
\newblock
\urldef\tempurl%
\url{https://arxiv.org/abs/2505.09388}
\showURL{%
\tempurl}


\bibitem[\protect\citeauthoryear{Zhang, Li, Cui, Cai, Liu, Fu, Huang, Zhao, Zhang, Chen, Wang, Luu, Bi, Shi, and Shi}{Zhang et~al\mbox{.}}{2023}]%
        {zhang2023sirens}
\bibfield{author}{\bibinfo{person}{Yue Zhang}, \bibinfo{person}{Yafu Li}, \bibinfo{person}{Leyang Cui}, \bibinfo{person}{Deng Cai}, \bibinfo{person}{Lemao Liu}, \bibinfo{person}{Tingchen Fu}, \bibinfo{person}{Xinting Huang}, \bibinfo{person}{Enbo Zhao}, \bibinfo{person}{Yu Zhang}, \bibinfo{person}{Yulong Chen}, \bibinfo{person}{Longyue Wang}, \bibinfo{person}{Anh~Tuan Luu}, \bibinfo{person}{Wei Bi}, \bibinfo{person}{Freda Shi}, {and} \bibinfo{person}{Shuming Shi}.} \bibinfo{year}{2023}\natexlab{}.
\newblock \showarticletitle{Siren's Song in the AI Ocean: A Survey on Hallucination in Large Language Models}.
\newblock \bibinfo{journal}{\emph{CoRR}}  \bibinfo{volume}{abs/2309.01219} (\bibinfo{date}{Sept.} \bibinfo{year}{2023}).
\newblock
\urldef\tempurl%
\url{https://arxiv.org/abs/2309.01219}
\showURL{%
\tempurl}


\bibitem[\protect\citeauthoryear{Zhao, Li, Zhang, Huang, Zhang, Chen, Shi, Li, and Chen}{Zhao et~al\mbox{.}}{2025}]%
        {llmidxadvis}
\bibfield{author}{\bibinfo{person}{Xinxin Zhao}, \bibinfo{person}{Haoyang Li}, \bibinfo{person}{Jing Zhang}, \bibinfo{person}{Xinmei Huang}, \bibinfo{person}{Tieying Zhang}, \bibinfo{person}{Jianjun Chen}, \bibinfo{person}{Rui Shi}, \bibinfo{person}{Cuiping Li}, {and} \bibinfo{person}{Hong Chen}.} \bibinfo{year}{2025}\natexlab{}.
\newblock \showarticletitle{LLMIdxAdvis: Resource-Efficient Index Advisor Utilizing Large Language Model}.
\newblock \bibinfo{journal}{\emph{CoRR}}  \bibinfo{volume}{abs/2503.07884} (\bibinfo{date}{March} \bibinfo{year}{2025}).
\newblock
\urldef\tempurl%
\url{https://doi.org/10.48550/arXiv.2503.07884}
\showURL{%
\tempurl}


\bibitem[\protect\citeauthoryear{Zhao, Cong, Shi, and Miao}{Zhao et~al\mbox{.}}{2022}]%
        {queryformer}
\bibfield{author}{\bibinfo{person}{Yue Zhao}, \bibinfo{person}{Gao Cong}, \bibinfo{person}{Jiachen Shi}, {and} \bibinfo{person}{Chunyan Miao}.} \bibinfo{year}{2022}\natexlab{}.
\newblock \showarticletitle{QueryFormer: a tree transformer model for query plan representation}.
\newblock \bibinfo{journal}{\emph{Proc. VLDB Endow.}} \bibinfo{volume}{15}, \bibinfo{number}{8} (\bibinfo{date}{April} \bibinfo{year}{2022}), \bibinfo{pages}{1658–1670}.
\newblock
\showISSN{2150-8097}
\urldef\tempurl%
\url{https://doi.org/10.14778/3529337.3529349}
\showDOI{\tempurl}


\bibitem[\protect\citeauthoryear{Zhao, Li, and Cong}{Zhao et~al\mbox{.}}{2023}]%
        {qpeval}
\bibfield{author}{\bibinfo{person}{Yue Zhao}, \bibinfo{person}{Zhaodonghui Li}, {and} \bibinfo{person}{Gao Cong}.} \bibinfo{year}{2023}\natexlab{}.
\newblock \showarticletitle{A Comparative Study and Component Analysis of Query Plan Representation Techniques in ML4DB Studies}.
\newblock \bibinfo{journal}{\emph{Proc. VLDB Endow.}} \bibinfo{volume}{17}, \bibinfo{number}{4} (\bibinfo{date}{Dec.} \bibinfo{year}{2023}), \bibinfo{pages}{823–835}.
\newblock
\showISSN{2150-8097}
\urldef\tempurl%
\url{https://doi.org/10.14778/3636218.3636235}
\showDOI{\tempurl}


\bibitem[\protect\citeauthoryear{Zhou, Lin, Zhou, and Li}{Zhou et~al\mbox{.}}{2024}]%
        {breakitdown}
\bibfield{author}{\bibinfo{person}{Wei Zhou}, \bibinfo{person}{Chen Lin}, \bibinfo{person}{Xuanhe Zhou}, {and} \bibinfo{person}{Guoliang Li}.} \bibinfo{year}{2024}\natexlab{}.
\newblock \showarticletitle{Breaking It Down: An In-Depth Study of Index Advisors}.
\newblock \bibinfo{journal}{\emph{Proc. VLDB Endow.}} \bibinfo{volume}{17}, \bibinfo{number}{10} (\bibinfo{date}{June} \bibinfo{year}{2024}), \bibinfo{pages}{2405–2418}.
\newblock
\showISSN{2150-8097}
\urldef\tempurl%
\url{https://doi.org/10.14778/3675034.3675035}
\showDOI{\tempurl}


\bibitem[\protect\citeauthoryear{Zhou, Liu, Li, Jin, Li, Wang, and Feng}{Zhou et~al\mbox{.}}{2022}]%
        {autoindex2022}
\bibfield{author}{\bibinfo{person}{Xuanhe Zhou}, \bibinfo{person}{Luyang Liu}, \bibinfo{person}{Wenbo Li}, \bibinfo{person}{Lianyuan Jin}, \bibinfo{person}{Shifu Li}, \bibinfo{person}{Tianqing Wang}, {and} \bibinfo{person}{Jianhua Feng}.} \bibinfo{year}{2022}\natexlab{}.
\newblock \showarticletitle{AutoIndex: An Incremental Index Management System for Dynamic Workloads}. In \bibinfo{booktitle}{\emph{ICDE}}.
\newblock


\bibitem[\protect\citeauthoryear{Zhou, Muresanu, Han, Paster, Pitis, Chan, and Ba}{Zhou et~al\mbox{.}}{2023}]%
        {zhou2023large}
\bibfield{author}{\bibinfo{person}{Yongchao Zhou}, \bibinfo{person}{Andrei~Ioan Muresanu}, \bibinfo{person}{Ziwen Han}, \bibinfo{person}{Keiran Paster}, \bibinfo{person}{Silviu Pitis}, \bibinfo{person}{Harris Chan}, {and} \bibinfo{person}{Jimmy Ba}.} \bibinfo{year}{2023}\natexlab{}.
\newblock \showarticletitle{Large Language Models are Human-Level Prompt Engineers}. In \bibinfo{booktitle}{\emph{The Eleventh International Conference on Learning Representations}}.
\newblock
\urldef\tempurl%
\url{https://openreview.net/forum?id=92gvk82DE-}
\showURL{%
\tempurl}


\end{thebibliography}

\end{document}